\documentclass[10pt,aps,prd,notitlepage,superscriptaddress,nofootinbib,amsmath,amssymb,twocolumn,showpacs]{revtex4-1}
\usepackage[latin9]{inputenc}

\pdfoutput=1

\usepackage{color,marvosym}
\usepackage{graphicx} 
\usepackage{comment}
\usepackage{breakurl}

\definecolor{darkred}{rgb}{0.5,0,0}
\definecolor{darkgreen}{rgb}{0,0.5,0}
\definecolor{darkblue}{rgb}{0,0,0.5}

\usepackage{hyperref}
\hypersetup{colorlinks,
linkcolor=darkblue,
filecolor=darkgreen,
urlcolor=darkblue,
citecolor=darkblue }
\newcommand{\fett}[1]{\boldsymbol{#1}}
\newcommand{\dd}{{\rm{d}}}
\newcommand{\ii}{{\rm{i}}}

\makeatother

\begin{document}

\title{Quasilinear observables in dark energy cosmologies}

\author{Cornelius Rampf}
\email{rampf@thphys.uni-heidelberg.de}
\affiliation{Institute for Theoretical Physics, University of Heidelberg, Philosophenweg 16, D--69120 Heidelberg, Germany}
\affiliation{Department of Physics, Israel Institute of Technology -- Technion,
Haifa 32000, Israel}

\author{Eleonora Villa}
\email{evilla@sissa.it}
\affiliation{SISSA, via Bonomea 265, 34136, Trieste, Italy}
\affiliation{INFN, Sezione di Trieste, Via Valerio 2, 34127 Trieste, Italy}

\author{Luca Amendola}
\email{l.amendola@thphys.uni-heidelberg.de}
\affiliation{Institute for Theoretical Physics, University of Heidelberg, Philosophenweg 16, D--69120 Heidelberg, Germany}

\date{\today}

\begin{abstract} 
What are the fundamental limitations of reconstructing the properties of dark energy, given cosmological observations in the quasilinear regime in a range of redshifts, to be as precise as required? The aim of this paper is to address this question by constructing model-independent observables, while completely ignoring practical  problems of real-world observations. Non-Gaussianities already present in the initial conditions are not directly accessible from observations,  because of a perfect degeneracy with the non-Gaussianities arising from the (weakly) nonlinear matter evolution in generalized dark energy models. By imposing a specific set of evolution equations that should cover a range of dark energy cosmologies, we find, however, a constraint equation for the linear structure growth rate $f_1$ expressed in terms of model-independent observables. Entire classes of dark energy models which do not satisfy this constraint equation could be ruled out, and for models satisfying it we could reconstruct e.g.\ the nonlocal bias parameters $b_1$ and $b_2$.
\end{abstract}


\maketitle

%%%%%%%%%%%%%%%%%%%%%%%%%%%%%%%%%%%%%%%%%%%%%%%%%%%%
\section{Introduction}\label{sec:intro}

Gravity is a nonlinear phenomenon that is also responsible for today's observed 
large-scale structure.  On large scales, gravitational interactions should 
be close to linear,
even if there are  significant non-Gaussian features in the 
initial conditions for structure formation. The linearity of gravitational 
interactions on large scales is mainly due to a suppressed interaction rate of 
energy fluctuations close to the causality horizon. On smaller scales 
gravitational interactions grow exponentially and
we observe nonlinear amplifications of 
over- and underdensities, accompanied by increasing tidal interactions.
However, there should be an intermediate regime where linear theory provides
a reasonably good approximation of the underlying physics, and nonlinearities
can be viewed as a small perturbation to it. This is what we call 
the quasilinear regime, 
where we expect that a theory that includes the leading nonlinearities should
deliver better approximations as opposed to a strictly linear analysis.

Cosmological structures such as filaments, clusters and voids emerge on scales 
that connect also such intermediate scales.
Galaxies are tracers of the underlying matter distribution, and the explicit 
bias relation is unknown. Generally, galaxy bias could 
depend on the scale and on nonlocal physical processes, 
such as galaxy formation and hydrodynamical interactions, whose specific
 mechanisms are not yet comprehensively understood.
Simplified bias models such as the local model can be very accurate, 
especially on large scales,  but need to be revised when investigating 
cosmological models beyond $\Lambda$CDM. The reason for the necessary revision 
is that departures from $\Lambda$CDM usually imply scale-dependent matter growth, 
which also renders the bias to be scale dependent and nonlocal 
(see \cite{Desjacques:2016bnm} for a review).

Dark energy (DE) could affect all of the above. 
So far the simple $\Lambda$CDM model has been remarkably successful at explaining 
a host of astrophysical observations on a wealth of scales, but more sophisticated 
DE models are not ruled out and should be further 
investigated \cite{Amendola,Clifton:2011jh,Kunz:2012aw}.
Much effort has been made to understand DE and possible modifications
at the level of background, linear and weakly nonlinear perturbation observables, 
often with the premiss to fix a particular DE model and investigate the 
resulting phenomenological consequences (e.g.\ \cite{Li:2011qda,Brax:2011aw,Barreira:2012kk,Li:2012by,Jennings:2012pt,Joudaki:2012bcf,Barreira:2013jma,Bellini:2015wfa}).
In the literature there are also many 
approaches to investigate DE modifications in a model independent way \cite{Huterer:2002hy,Zhang:2007nk,Shafieloo:2009hi,Zhao:2009fn,Zhao:2010dz,Shafieloo:2011zv,Stebbins:2012vw,Valkenburg:2012td,Amendola:2012ky,Baker:2013hia,Alam:2015rsa,Taddei:2016iku}, 
but they are usually restricted to the linear regime.  
One of the tasks of the present study is to extend the model-independent 
approach by allowing weak nonlinearities in the analysis.

Quasilinear observables probe quasilinear scales, and on very large scales where
the physics is linear, quasilinear observables should deliver, to a very 
good approximation, the same answers as linear observables.
By also allowing weak nonlinearities in our model-independent analysis, 
we of course not only provide access to more scales,
but also introduce many more unknowns that should be 
taken into account. Such unknowns could arise from e.g.\ the bias model or 
the weakly nonlinear matter evolution within the DE model. Furthermore, also 
non-Gaussian modifications could be present already in the initial conditions 
of structure formation. These modifications are usually dubbed as primordial 
non-Gaussianity (PNG), and in the present paper we refrain to use any simplified 
parametrization of PNG. Rather we show, amongst other things, that 
PNG and non-Gaussianities arising from the weakly nonlinear matter evolution
are indistinguishable, because of a 
perfect degeneracy  in \mbox{DE models.}
However, by going beyond linear order we derive new observables that 
constrain the combined effect of gravity and PNG and, furthermore, find a 
novel constraint equation that also gives insight into the linear regime of 
structure formation --- much more insight than could be achieved in a strictly 
linear analysis.

In the present paper, which is closest in the spirit of Ref.~\cite{Amendola:2012ky},
we completely ignore practical problems of actual observations, such as survey 
geometry and we assume good-enough statistics. Thus, we investigate a vastly idealized 
scenario with the aim of obtaining the fundamental limitations of reconstructing the 
properties of DE cosmologies.

This paper is organized as follows. In the following section we outline the 
theoretical assumptions and approximations that we use in this paper. 
In Sec.\,\ref{sec:PTmain} we apply our assumptions and approximations and develop
the weakly nonlinear framework for generalized DE models.
Sections~\ref{sec:power}--\ref{sec:Lens-GaBisp} introduce various statistical 
estimators that are used to connect the theory with galaxy and weak gravitational lensing observations.
Readers who like to skip the technical details should at least read the short 
Sec.\,\ref{sec:power} where we explain our methodology that we apply throughout this paper.
Then, in Sec.\,\ref{sec:Observables} we report a selection of quasilinear observables from the statistical estimators (see Appendix~\ref{app:moreObs} for a complete list of observables).
We derive equations that deliver model-independent constraints of several unknowns in Sec.\,\ref{sec:consistency}.
Our constraint equations, although being based on fairly general 
assumptions, still rely on a given class of theoretical models 
that should hold for many DE cosmologies in a suitable range of cosmological scales. 
We do not rule out the possibility that theoretical improvements could allow to 
extend the validity regime of our analysis,
and in Sec.\,\ref{sec:valid}  we sketch a few of such theoretical avenues.
Finally, we conclude in~\ref{sec:concl}.

We adopt metric signature \mbox{$(-++\,+)$,} 
cosmic time is $t$ and its corresponding partial derivative is the overdot,  
while a prime denotes a partial derivative with respect to the time variable $N=\ln a$, 
where $a=(1+z)^{-1}$ is the cosmic scale factor and $z$ the redshift. 
The subscript $0$ denotes present time. 
If not otherwise stated, the functional dependence in Fourier space 
is with respect to $k= |\fett{k}|$.
The shorthands $\fett{k}_{12}$ and $\fett{k}_{123}$ stand for 
$\fett{k}_{1}+\fett{k}_{2}$ and $\fett{k}_{1}+\fett{k}_{2}+\fett{k}_{3}$, respectively,
and we make use of the integral shorthand notation 
$\int \dd^3 \fett{k}_{12} = \int \dd^3 \fett{k}_1 \int \dd^3 \fett{k}_2$.
For a given function ${\cal F}(\fett{k}_i,\fett{k}_j)$
that depends on two wave vectors $\fett{k}_i$ and $\fett{k}_j$  where 
$i, j \ni \{ 1,2,3\}$ and $i \neq j$,
 the shorthand ${\cal F}^{\rm eq}$ denotes 
equilateral dependence for which
$k_1 = k_2 = k_3  \equiv k$. We apply a similar shorthand
 ${\cal F}^{\rm sq}_{ij} \equiv {\cal F}^{\rm sq} (\fett{k}_i,\fett{k}_j)$ 
for triangle dependences in the squeezed limit, 
where $k_1 = k_2 \equiv k$ and $k_3 = \Delta k$, and $\Delta k/k \to 0$. 
 
\newpage

%%%%%%%%%%%%%%%%%%%%%%%%%%%%%%%%%%%%%%%%%%%%%%%%%%%%%%%
\section{Assumptions \&  Approximations}\label{sec:setup}

In the present work
the considered departures from $\Lambda$CDM are described by 
two free functions, the first being a modification of the source term in the Poisson equation (usually dubbed $Y(z;k)$), and the second being a modification of the
gravitational lensing potential (often called $\Sigma(z;k)$). 
Such deviations occur for example in modified theories of gravity or 
coupled DE models (see e.g.\ \cite{Amendola}).
Regarding the asssumptions on the underlying geometry and matter content of the 
Universe, we impose that: 
\begin{itemize}
  \item[(a)] The background geometry of the Universe is well described by a
    Friedmann--Lema\^itre--Robertson--Walker metric; 
    its evolution is parametrized by the cosmic scale factor $a(t)$. 
    The Hubble parameter $H= \dot a /a$ is governed by the Friedmann equation 
    \begin{equation} \label{friedmann}
      H^2 - H_0^2 \Omega_{\rm k0} a^{-2} = 
         \frac 1 3 \left( \bar \rho_{\rm m} + \bar \rho_{\rm x}\right) 
    \end{equation}
    (setting $8\pi G =1$), where $H_0$ and $\Omega_{\rm k0}$ are, respectively, 
    the present day values of the Hubble parameter and curvature, 
    $\bar \rho_{\rm m} \sim a^{-3}$ is the background density of matter, 
    and $\bar \rho_{\rm x}$ is the combined background density of an unspecified 
    modification of gravity. 
    Background observations can generally measure $H(z)$ up to a multiplicative 
    constant (see e.g., \cite{Amendola:2012ky}), and, thus, we assume in the following 
    that the dimensionless Hubble function ${\cal E}(z) \equiv H(z)/H_0$ is an 
    observable. Combining measurements of the luminosity or angular-diameter distance 
    with $H(z)$, we can furthermore determine $\Omega_{\rm k0}$ 
    \cite{Netterfield:2001yq}. By contrast, it is impossible to measure 
    $\Omega_{\rm m0}$ without invoking an explicit parametrization 
    for $\bar \rho_{\rm x}$, as the problem is perfectly degenerated \cite{Kunz:2007rk}.
 
  \item[(b)] The matter content (i.e., dark matter and baryons) is
    exposed to an identical gravitational force and moves on geodesics 
    described by a given metric theory.
    This assumption in particular 
    restricts our approach to sufficiently large scales where baryonic 
    feedback is negligible. 
    For example, in a $\Lambda$CDM universe, baryons affect the 
    velocity divergence power spectrum of dark matter by less than 1\% 
    at scales larger than $0.5\,h/$Mpc \cite{Hellwing:2016ucy}.

    Probing very large scales requires a careful assessment of so-called 
    secondary effects that naturally arise in metric theories of gravity.
    For example, in $\Lambda$CDM which is based on general relativity, 
    such secondary effects are relativistic  corrections that 
    appear at the matter level \cite{Villa:2015ppa}, 
    through radiation \cite{Tram:2016cpy}, or 
    through light-cone effects~\cite{DiDio:2016gpd}.
    It is beyond the scope of this paper to incorporate such effects, and we, thus,
    restrict our analysis to the subhorizon regime to minimize its contamination.
    Relativistic corrections in the initial conditions, however, could generate 
    nonzero intrinsic bispectra,  
    which we do allow in our analysis; \mbox{see (e).}

   \item[(c)] We are interested in the cosmological evolution of a single-stream 
    matter fluid within the quasilinear regime where
    perturbation theory should give meaningful results. 
    We only take the 
    leading nonlinearities into account and consequently ignore any loop 
    contributions. This implies that we only need to go up to second order in the 
    fluid variables, i.e., 
    \begin{equation} \label{PTansatz}
      \delta_{\rm m} = \delta_{\rm m1} + \delta_{\rm m2} \,, \qquad \theta_{\rm m} 
          =  \theta_{\rm m1} + \theta_{\rm m2} \,,
    \end{equation}
    where $\delta_{\rm m} \equiv (\rho_{\rm m} - \bar \rho_{\rm m})/ \bar \rho_{\rm m}$ 
    is the matter density contrast and 
    $\theta_{\rm m} \equiv \nabla \cdot \fett{v}_{\rm m}$ the  
    divergence of a rescaled peculiar velocity $\fett{v}_{\rm m} \equiv \fett{v}_{\rm m, pec}/(a H)$.
    In the following section we provide some evolution equations for these fluid 
    variables, although explicit evolution equations are only 
    required for Sec.\,\ref{sec:consistency} when we derive a novel 
    constraint equation. 

    The theoretical tools used in this paper are based on standard 
    perturbation theory (SPT) \cite{Bernardeau:2001qr}. 
    Several theoretical models in the literature exist that could push the validity 
    of the perturbative description to more nonlinear scales. We
    defer the discussion about such avenues to Sec.\,\ref{sec:valid}.

  \item[(d)] We apply the so-called plane-parallel limit when projecting fluid 
    variables from real-space coordinates $\fett{x}$ to redshift-space 
    coordinates $\fett{s}$ \cite{Kaiser:1987qv}, 
    \begin{equation}
      \fett{s} = \fett{x} + \nabla^{-2} \nabla_z \theta \,, 
    \end{equation}
    where the inverse Laplacian is with respect to the real-space coordinates.
    For the bias relation between matter and galaxy, we allow the bias function 
    to be scale and time dependent. This means that the galaxy density can be 
    written as $\delta_{\rm g} = \delta_{\rm g1} + \delta_{\rm g2}$, which is in 
    Fourier space \cite{Matsubara:2011ck}
    \begin{subequations}
      \begin{align} 
        \,\qquad\delta_{\rm g1}(z;\fett{k}) &= b_1 \delta_{\rm m1} \,, \label{deltag1} \\ 
        \,\qquad\delta_{\rm g2}(z;\fett{k}) &= b_1 \delta_{\rm m2} + \frac 1 2 \int \frac{\dd^3 \fett{k}'}{(2\pi)^3} b_2(\fett{k}', \fett{k}-\fett{k}') \nonumber \\ &\,\qquad\hspace{1.24cm}\times \delta_{\rm m1}(\fett{k}')\,\delta_{\rm m1}(\fett{k}-\fett{k}')  \,, \label{deltag2}
      \end{align}
    \end{subequations}
    where, again, the unknown bias functions $b_1$ and $b_2$ are generally 
    scale and time dependent.
    By virtue of assumption~(b), we assume that there is no bias between the matter 
    and galaxy velocity, i.e., \mbox{$\fett{v}_{\rm m} = \fett{v}_{\rm g}$.}

  \item[(e)] Although there is currently no sign of any significant nonzero primordial
    non-Gaussianity  (PNG),  we allow in the present analysis for the most general 
    deviation of Gaussian initial conditions, that is we {\it do  not invoke} any 
    parametrization of PNG. Rather we assume that there exists a possible nonzero 
    intrinsic matter bispectrum,
    \begin{equation}
      \left\langle \delta_{\rm m1} \delta_{\rm m1} \delta_{\rm m1} \right\rangle_{\rm c}
          \sim  B_{\rm m111} (z;\fett{k}_1, \fett{k}_2, \fett{k}_3) \,, 
    \end{equation}
    which is related in some arbitrary way to the initial curvature perturbation on 
    superhorizon scales.
    For example, $B_{\rm m111}$ could originate from PNG of the local type in which 
    case one would expect large contributions to the bispectrum in the squeezed limit. 
    We neglect PNG contributions to the initial trispectrum and higher-order 
    correlators, as they usually involve loop corrections; see assumption (c).

\end{itemize}

%%%%%%%%%%%%%%%%%%%%%%%%%%%%%%%%%%%%%%%%%%%%%%%%%%%%%%%
\section{\!Equations in real \& redshift space\!} \label{sec:PTmain} 

Let us apply the above assumptions and approximations, 
and set up the respective
equations, firstly in real space and then redshift space. 
Although not required for the bulk part of this paper 
(except Sec.\,\ref{sec:consistency}), let us assume the following {\it explicit} 
fluid equations for matter,
\begin{subequations}  \label{evoeqs}
\begin{align}
 \delta_{\rm m}' + \nabla \cdot \left( 1 + \delta_{\rm m} \right) \fett{v}_{\rm m}  
     &= 0 \,, \label{eq:conti} \\
\fett{v}_{\rm m}' + \left( \fett{v}_{\rm m} \cdot \nabla \right) \fett{v}_{\rm m}
     &= -  \left( 2+\frac{H'}{H} \right) \fett{v}_{\rm m} - \nabla \Psi \,. \label{eq:mom}
\end{align}
Here $\delta_{\rm m} \equiv (\rho_{\rm m} - \bar \rho_{\rm m})/ \bar \rho_{\rm m}$ 
is the matter density contrast and 
$\fett{v}_{\rm m} \equiv \fett{v}_{\rm m,pec}/(aH)$ the rescaled 
peculiar velocity of matter,
$H$ the Hubble parameter, a prime denotes a partial derivative with respect to the
time variable $N=\ln a$, and $a$ is the cosmic scale factor itself determined by 
the Friedmann equation~(\ref{friedmann}). 
We make use of the modified Poisson equation,
\begin{align}
  \nabla^{2}\Psi(\fett{x})=\frac{3}{2}\Omega_{{\rm m}} \int \dd^3 \fett{y} \,Y(\fett{x}-\fett{y})\,\delta_{\rm m}(\fett{y})  \,, \label{eq:modPois}
\end{align}
\end{subequations}
where $Y$ is a scale- and time-dependent clustering function. 
The function $Y$ is by definition equal to unity in a universe 
that is prescribed by a cosmological constant ($\Lambda$) and a matter component; 
by contrast, for a {\it realistic} $\Lambda$CDM universe, 
where generally not just matter but also other fluid components are present 
(e.g., massive neutrinos), $Y$ differs (mildly) from unity reflecting 
the fact  that matter couples to other fluid components gravitationally.
In addition, $Y\neq 1$ can be established by a wealth of modified gravity scenarios 
(see~\cite{Amendola} and references therein). In the following, we will make 
no model-dependent assumptions how $Y$ might look alike and thus leave
it as a free function.

To solve Eqs.\,(\ref{eq:conti})--(\ref{eq:modPois}), we assume that the fluid motion 
is irrotational and thus, the velocity can be fully described by its 
divergence, $\theta_{\rm m} = \nabla \cdot \fett{v}_{\rm m}$. Perturbing the density and 
velocity according to~(\ref{PTansatz}), we obtain to first order in Fourier space 
\begin{equation}
  \delta_{\rm m1}''+\left(2+\frac{H'}{H}\right) \delta_{\rm m1}' - \frac{3}{2}\Omega_{{\rm m}}Y \delta_{\rm m1} = 0 \,. \label{evoLinearDelta}
\end{equation}
The growing mode solution for the density can be formally written as
$\delta_{\rm m1}(z;k)= D(z;k) \,\delta_{0}(z_0;k)$, where $D$ is the linear growth 
function which is normalized to unity today, and $\delta_0$ is the present matter 
density. 
We note that $D$ is not only time dependent but %in many modified gravity scenarios 
in general also scale dependent.
Using the solution for the density, one immediately gets for the first-order velocity 
$\theta_{\rm m1} = - \delta_{\rm m1}'= - f_1 \delta_{\rm m1}$,  
where the linear structure growth rate $f_{1}$ is defined by $f_{1}\equiv D'/D$. 

Second-order solutions can be formally written as
\begin{subequations} \label{pertAnsatz}
\begin{align}
 \delta_{\rm m2}(z; k) & = \int\frac{\dd^{3}\fett{k}_{12}}{(2\pi)^{3}} \delta_{{\rm D}}^{(3)}\left(\fett{k}-\fett{k}_{12} \right) {\mathfrak{F}}_{2}(z; \fett{k}_{1}, \fett{k}_2 ) \nonumber  \\ 
    &\qquad \qquad \hspace{1cm}\times \delta_{\rm m1}(z;{k}_{1})\, \delta_{\rm m1}(z; {k}_{2})\,,  \\  
 \theta_{\rm m2}(z; k) &= \int\frac{\dd^{3}\fett{k}_{12}}{(2\pi)^{3}} \delta_{{\rm D}}^{(3)}\left(\fett{k}-\fett{k}_{12} \right) {\mathfrak{G}}_{2}(z; \fett{k}_{1}, \fett{k}_2 )  \nonumber \\ 
    &\qquad \qquad \hspace{1cm}\times \delta_{\rm m1}(z;{k}_{1})\, \delta_{\rm m1}(z; {k}_{2})\,,  
\end{align}
\end{subequations}
where $\delta_{{\rm D}}^{(3)}$ is the Dirac-delta distribution,  ${\mathfrak{F}}_2$ and ${\mathfrak{G}}_2$ are perturbation kernels with symmetric $\fett{k}$-dependence in their arguments,
and the matter density and velocity only depend on the magnitude of the
wave vector $k \equiv |\fett{k}|$, due to statistical isotropy. 
For an Einstein-de Sitter (EdS) universe the above kernels become time independent, and read in our sign convention 
\begin{align}
{\mathfrak{F}}_{2}^{\rm EdS}  &= \frac 5 7 + \frac{\fett{k}_{1}\cdot \fett{k}_{2}}{2k_{1}k_{2}} \left[\frac{k_{1}}{k_{2}}+ \frac{k_{2}}{k_{1}} \right]+ \frac 2 7 \left( \frac{\fett{k}_{1}\cdot\fett{k}_{2}}{k_{1}k_{2}} \right)^{\!\!2} ,  \\
{\mathfrak{G}}_{2}^{\rm EdS} &=- \frac 3 7 - \frac{\fett{k}_{1}\cdot\fett{k}_{2}}{2k_{1}k_{2}} \left[ \frac{k_{1}}{k_{2}}+ \frac{k_{2}}{k_{1}} \right]- \frac 4 7 \left( \frac{\fett{k}_{1}\cdot\fett{k}_{2}}{k_{1}k_{2}} \right)^{\!\!2} \!. 
\end{align}
For an EdS universe these kernels are well known in the 
literature \cite{Bernardeau:2001qr}, 
and are usually labelled with $F_2$ and $G_2$, respectively.
For a standard $\Lambda$CDM model or in modified gravity, 
however, these kernels generally do depend on time and could have a more 
complicated $k$-dependence.
Note that to construct the observables in the following, 
we do not require explicit solutions for 
${\mathfrak{F}}_2$ and ${\mathfrak{G}}_2$; we only assume that solutions for $\delta_{\rm m}$ and $\theta_{\rm m}$
can be written in terms of a power series in the linear density.
Such perturbative solutions should describe the physics sufficiently well, 
provided that nonlinear corrections are small with respect to the linear contributions, and that on the considered scales vorticities and the effects from velocity dispersion can be neglected (see Sec.\,\ref{sec:valid} for details).
Furthermore, since in this paper we consider the modification $Y$ of the source term in the Poisson equation as a free function, also ${\mathfrak{F}}_2$ and ${\mathfrak{G}}_2$ are effectively free functions since they depend \mbox{on $Y$.}
% (note however that we do require that ${\mathfrak{F}}_2$ vanishes for certain configurations in the so-called squezzed limit, see Appendix~\ref{app:squezzed} for a derivation).}

Up to this point we have dealt with matter perturbations in real space,
but what we observe are galaxies, measured in redshift space.
We deal with the galaxy description as outlined in Sec.\,\ref{sec:setup}, see in 
particular Eqs.\,(\ref{deltag1})--(\ref{deltag2}), where we employ 
a nonlocal bias description between the matter density $\delta_{\rm m}$ and
the galaxy density $\delta_{\rm g}$, 
whereas according to assumption (d) we assume
$\fett{v}_{\rm m} = \fett{v}_{\rm g} \equiv \fett{v}$.
The next step is to incorporate the effects of redshift-space distortions,
resulting from the fact that the observed comoving positions
of galaxies $\fett{s}$ are modified by their peculiar motion according 
to $\fett{s}=\fett{x}+v_{z}(\fett{x})\fett{\hat{z}}$ in the plane-parallel limit, 
where  $ \hat{\fett z}$ is the unit vector along the line of sight and $v_z$ is the 
projection of the peculiar velocity along the $z$ axis.
This leads to the following relation between the galaxy density in redshift space, $\delta_{{\rm g}}^{{\rm s}}(z;\fett{k})$, and the one in real space, \mbox{$\delta_{{\rm g}}(z;\fett{x})$, i.e., \cite{Matsubara:2011ck}}
\begin{align}
 \delta_{{\rm g}}^{{\rm s}}(z;\fett{k}) &= \int\dd^{3}\fett{x}\,{\rm e}^{-\ii\fett{k}\cdot\fett{x}} \left[ 1 + \delta_{{\rm g}}(z;\fett{x}) \right]{\rm e}^{-\ii k_z v_{z}} \,.  
\intertext{Taylor expanding the fluid variables and the exponential in the last expression, we obtain}
 \delta_{{\rm g1}}^{{\rm s}}(z;\fett{k}) &= S_1(z;\fett{k}_1)\, \delta_{\rm m1}(z;\fett{k}_{1}) \,,  \\
 \delta_{{\rm g2}}^{{\rm s}}(z;\fett{k}) &= \int\frac{\dd^{3}\fett{k}_{12}}{(2\pi)^{3}} \delta_{{\rm D}}^{(3)}(\fett{k} - \fett{k}_{12})\,S_{2}(z;\fett{k}_{1},\fett{k}_2)  \nonumber  \\ 
  &\qquad \hspace{1.53cm} \times \delta_{\rm m1}(z;\fett{k}_{1}) \,\delta_{\rm m1}(z;\fett{k}_{2})\,, 
\end{align}
with the kernels
\begin{align}
S_{1} &= b_{1}+f_{1} \mu_{1}^{2} \,, \\  
S_{2} &=-\mu_{12}^{2} \,{\mathfrak{G}}_{2} + b_{1}(k_{12}) {\mathfrak{F}}_{2} \nonumber \\ 
  &+\frac 1 2 \mu_{12}k_{12}\left[\frac{\mu_{1}}{k_{1}}b_{1}(k_{2})f_{1}(k_{1})
     +\frac{\mu_{2}}{k_{2}}b_{1}(k_{1})f_{1}(k_{2})\right]  \nonumber \\
  & +\frac{(\mu_{12}k_{12})^{2}}{2}\frac{\mu_{1}\mu_{2}}{k_{1}k_{2}}f_{1}(k_{1})f_{1}(k_{2})+\frac{1}{2}b_{2}(\fett{k}_{1},\fett{k}_{2}) \,, 
\end{align}
where $\mu=\fett{k}\cdot\hat{\fett{z}}/k$ is the cosine of the angle
formed by the direction of the observation $\hat{\fett{z}}$ and the
wave vector $\fett{k}$ and $\mu_{i}=\fett{k}_{i}\cdot\hat{\fett{z}}/k_{i}$. 
The kernel $S_1$ is widely known in the literature,
in particular also in the frameworks of nonlocal bias and DE models.
To our knowledge, the second-order kernel $S_2$ has not been reported earlier in the context of DE models.
In the framework of nonlocal bias a very similar kernel, however valid only for a $\Lambda$CDM universe, has been derived in Ref.\,\cite{Matsubara:2011ck}, and in the respective limit our $S_2$ agrees with the one of~\cite{Matsubara:2011ck}.

%%%%%%%%%%%%%%%%%%%%%%%%%%%%%%%%%%%%%%%%%%%%%%%%%%%%%%%%%%%
\section{Power spectrum in redshift space}\label{sec:power}

To understand our methodology in the following sections, it is instructive to 
first investigate the linear observables that can be constructed from the 
galaxy power spectrum  \cite{Amendola:2012ky}. 
The %equal-time 
galaxy power spectrum in redshift space is defined as
\begin{equation}
\left\langle\delta_{\rm g}^{\rm s}(\fett{k}_{1})\delta_{\rm g}^{\rm s}(\fett{k}_{2}) \right\rangle_{\rm c} = (2\pi)^{3}\delta_{\rm D}^{(3)}(\fett{k}_{12}) P_{\rm g}^{\rm s}(z; \mu_1, k_{1}) \,,  
\end{equation}
where we note that $P_{\rm g}^{\rm s}$ depends not only on the magnitude but also 
on the cosine of the wave vector with respect to the direction of the observation, since it 
acquires an angular dependence due to the redshift-space distortions. The matter 
power spectrum $P_{\rm m}(k)$, by contrast, depends only on the modulus $k$ due to 
the assumption of statistical isotropy.

As for the perturbations of field variables, we can formulate the power spectrum 
in terms of a power series within perturbation theory, e.g., for the matter power 
spectrum we  have, to the leading order, 
that $P_{\rm m} = P_{\rm m11} \sim \delta_{\rm m1}^2$.
In the linear regime and for scales much smaller than the survey characteristic size, 
one can write for the galaxy power spectrum
\begin{equation} 
 P_{\rm g}^{\rm s}(z; k,\mu)= \left( b_1 + f_1 \mu^{2} \right)^{2} P_{\rm m11} \,, 
\end{equation}
where we remind the reader that the 
functions $f_1$ and $b_1$ depend generally on space and time.
This expression can be written in terms of a polynomial in $\mu$:
\begin{equation}
 P_{\rm g}^{\rm s}(z;k,\mu) = P_{\rm m11}(z;k) \sum_iP_{i} \,\mu^{i} \,, \label{eq:pg}
\end{equation}
with the only nonvanishing coefficients
\begin{equation}
 P_0  =  b_{1}^{2} \,, \qquad 
 P_2  =  2b_{1}f_{1} \,, \qquad
 P_4  =  f_{1}^{2} \,. \label{powercoeff}
\end{equation}
Observations can be made in principle at all values of~$\mu$. This
means that one can measure individually each term in the $\mu$ expansion.
Taking ratios of the various terms in Eq.\,(\ref{eq:pg}) one gets
rid of $P_{\rm m11}$ (and the unknown normalization $\sigma_8$), 
whose shape depends in general on initial conditions.
One obtains, for example, the quantity 
\begin{equation}
 {\cal P}_1 = f_{1}/b_{1}
\end{equation}
from $(2P_4)/P_2$.
The same procedure, extended
to the bispectrum, is at the core of the method presented below. In
addition to galaxy spectra, we will take into account also shear lensing
spectra and cross-correlation spectra of lensing and galaxy
clustering, in order to identify which quantities can be measured
directly from observations without assumptions on the shape of the
\mbox{(bi-)spectra.} 

Further linear observables are reviewed in Sec.\,\ref{sec:linearO}.

%%%%%%%%%%%%%%%%%%%%%%%
\section{Bispectrum in redshift space} \label{sec:GalaxyBisp}

We now continue with the next-to-leading order statistical estimator.
The %equal-time 
galaxy bispectrum in redshift space is defined as
\begin{align}
 \left\langle\delta_{\rm g}^{\rm s}(\boldsymbol{k}_{1})\delta_{\rm g}^{\rm s}(\boldsymbol{k}_{2}) \delta_{\rm g}^{\rm s}(\boldsymbol{k}_{3}) \right\rangle_{\rm c} = (2\pi)^{3}\delta_{\rm D}^{(3)}(\fett{k}_{123}) B_{{\rm g}}(\fett{k}_{1},\fett{k}_{2},\fett{k}_{3}).
\end{align}
The density bispectrum is nonzero only when 
non-Gaussianities in the density are present. 
This is especially 
the case in the quasilinear regime of structure formation, 
which encompasses also the linear regime.
As mentioned in Sec.\,\ref{sec:setup}, we allow in the present analysis of 
nonlinearities arising from the initial condition (PNG), and of 
the weakly nonlinear evolution of matter.

For the galaxy bispectrum, we get to the leading order
\begin{align} \label{generalbispec}
 B_{{\rm g}} &=2S_{2}(\fett{k}_{1},\fett{k}_{2})S_{1}(\fett{k}_{1})S_{1}(\fett{k}_{2})P_{{\rm m11}}(k_{1})P_{{\rm m11}}(k_{2}) \nonumber \\ 
 &+\text{two~perms} 
  + S_{1}(\fett{k}_{1})S_{1}(\fett{k}_{2}) S_{1}(\fett{k}_{3}) B_{\rm m111} \,,
\end{align}
where 
\begin{equation}  
  \left\langle \delta_{\rm m1}(\fett{k}_1) \delta_{\rm m1}(\fett{k}_2) \delta_{\rm m1}(\fett{k}_3) \right\rangle_{\rm c} = (2\pi)^3 \delta_{\rm D}^{(3)} (\fett{k}_{123}) B_{\rm m111}
\end{equation}
is the said non-Gaussian component arising from primordial/unknown physics.

The galaxy bispectrum in redshift space %in Eq.~(\ref{generalbispec}) 
is a function of five variables. The shape of the triangle is defined 
by three variables:  the length of two sides, i.e., the magnitude of 
two wave vectors, $k_1$ and $k_2$, and the angle between them, 
$\cos\theta_{12}= \fett{k}_{1}\cdot\fett{k}_{2}/(k_{1}k_{2})$. The two remaining 
variables characterize the orientation of the triangle with respect to 
the line of sight: we take them to be the polar angle 
of $\fett{k}_1$, $\omega= \arccos \mu_1$, and the azimuthal angle $\phi$ 
around $\fett{k}_1$. 
All the angles between the wave vectors and the line of sight can be written 
in terms of $\mu_1$ and $\phi$ \cite{Scoccimarro:1999ed},
\begin{align} \label{scocci}
\mu_1 &=  \frac{\fett{k}_{1}\cdot\hat{\fett{z}}}{k_{1}}\,, \quad \mu_2 = \mu_1 \cos \theta_{12} - \sqrt{1- \mu_1^2}\sin \theta_{12}\cos\phi\,, \nonumber \\
 \mu_3 &= - \frac{k_1}{k_3}\mu_1 -\frac{k_2}{k_3}\mu_2 \,.
\end{align}
We now determine the explicit expressions for the galaxy bispectrum 
for two fixed triangle configurations, namely for the equilateral and the squeezed type.

\subsection{The equilateral bispectrum}

In the equilateral configuration all the wave vectors have the same 
magnitude which we take to be 
$k_1 = k_2 = k_3 \equiv k$, from which it follows 
that $\fett{k}_i \cdot \fett{k}_j/(k_i k_j) =-1/2$, for $i \neq j$. 
Furthermore, the relation~(\ref{scocci}) between the three $\mu_i$'s simplifies to
\begin{align}
 \mu_{2} & =- \frac{\mu_{1}}{2} -\sqrt{3-3\mu_{1}^{2}} \frac{\cos\phi}{2}\,, \quad 
 \mu_{3}  =-\mu_{1}-\mu_2\,, \label{eq:cosinesEqui}
\end{align}
which we use to replace all $\mu_2$'s and $\mu_3$'s in the general expression for 
the bispectrum~(\ref{generalbispec}) in terms of $\mu_1$. 
We are thus left with a bispectrum that depends only on two angles, 
namely on $\mu_1$ and on the azimuthal angle $\phi$. We integrate out the 
azimuthal angle because of statistical isotropy around the redshift axis. 
Thus, one finally arrives at the equilateral 
bispectrum which is given in terms of a polynomial in $\mu_1$, 
\begin{equation} 
 B_{{\rm g}}^{{\rm eq}}= P_{{\rm m11}}^{2} \sum_{i}B_{i}^{{\rm eq}}\,\mu_{1}^{i}\,,
\end{equation}
with nonvanishing coefficients $B_0^{{\rm eq}}$, $B_2^{{\rm eq}}$, $B_4^{{\rm eq}}$,
$B_6^{{\rm eq}}$ and $B_8^{{\rm eq}}$. In the main text we only need the last two 
coefficients
\begin{align}
 B_6^{{\rm eq}} &=-\frac{177}{1024}  f_1^2  \left( f_1^2 + 16 {\mathfrak{G}}_2^{\rm eq} -\frac{8}{3}Q_{\rm m111}^{\rm eq} f_1 \right) \,, \\
 B_8^{{\rm eq}} &= -\frac{87}{1024}f_1^4 \,,
\end{align}
where ${\mathfrak{G}}_2^{\rm eq}$ is the second-order velocity kernel in the equilateral configuration, and we have defined the reduced intrinsic bispectrum
\begin{equation}
  Q_{\rm m111}^{\rm eq} \equiv B_{\rm m111}^{\rm eq} /P_{\rm m11}^2 \,.
\end{equation}
The complete list of bispectrum coefficients is given in Appendix~\ref{app:bispec}.

\subsection{The squeezed bispectrum}

The squeezed bispectrum is a specific limit that correlates density perturbations 
on essentially two different scales to each other. In that limit, two density 
perturbations which are usually taken to be well inside the horizon, are 
correlated with another perturbation close to the horizon (or beyond).
The corresponding triangle configuration in that limit is such that 
one wave vector, $\Delta k$, is much smaller
than the other two. 
We choose $k_{1}=k_{2} = k$, and $k_{3}=\Delta k$. We leave $\Delta k$ as a free
parameter but note that the squeezed approximation becomes more accurate when
$\Delta k/k \to 0$.
In the present paper we assume that the correlation length $k$ is
in the linear or in the quasilinear regime,
where second-order perturbation theory 
is a good approximation of the underlying physics, whereas $\Delta k$ is on 
sufficiently large scales where perturbations should mostly follow the overall 
Hubble flow and are otherwise well described by linear perturbation theory. 
For the squeezed bispectrum, 
we thus assume the existence of an intermediate regime where we can 
use the linear observables  as linear operators
on functions which depend on the squezzed bispectrum triangle 
side $\Delta k$ (see the following).

From the $\mu_i$ relations~\eqref{scocci}, we get $\mu_2 \simeq -\mu_1$ 
for all values of the azimuthal angle $\phi$, and the latter drops out. 
Thus, we can write the squeezed bispectrum as a polynomial of two cosines, 
\begin{align}
 B_{{\rm g}}^{{\rm sq}}=\sum_{i,j}B_{ij}^{{\rm sq}}\mu_{1}^{i}\mu_{\Delta k}^{j}\,,
\end{align}
with the only nonvanishing coefficients 
\begin{align}
 B_{00}^{\rm sq} &= 
   a_1  b_{1} b_{1,\Delta k} + b_{2,12}^{\rm sq} b_{1}^{2} P_{{\rm m11}}^{2}
   + B_{\rm m111}^{\rm sq} b_1^2 b_{1, \Delta k}  \,,   \\
\bar B_{02}^{\rm sq} &=  a_1  b_{1} b_{1,\Delta k} 
   + B_{\rm m111}^{\rm sq} b_1^2 b_{1,\Delta k} \,, \\
B_{20}^{\rm sq} &=  a_1  f_1  b_{1, \Delta k} + a_2 b_1 b_{1, \Delta k} 
   + 2 b_{2,12}^{\rm sq} b_{1} f_{1} P_{{\rm m11}}^{2} \nonumber \\
       &+ 2 B_{\rm m111}^{\rm sq} f_1 b_1 b_{1,\Delta k}   \,, \\
\bar B_{22}^{\rm sq} &=  a_1 f_1 b_{1, \Delta k} + a_2 b_1 b_{1, \Delta k}
         + 2 B_{\rm m111}^{\rm sq}  f_1 b_1 b_{1,\Delta k}   \,, \\
B_{40}^{\rm sq} &=  a_2 f_1 b_{1, \Delta k} +  b_{2,12}^{\rm sq} f_{1}^{2} P_{{\rm m11}}^{2} 
   + B_{\rm m111}^{\rm sq} f_1^2 b_{1,\Delta k}   \,, \\
 \bar B_{42}^{\rm sq} &=  a_2 f_1 b_{1, \Delta k} 
   + B_{\rm m111}^{\rm sq} f_1^2 b_{1,\Delta k} \,, 
\end{align}
where we have introduced the shorthand notation 
$b_{2,12}^{{\rm sq}}\equiv b_{2}^{{\rm sq}}(\fett{k}_{1},\fett{k}_{2})$,
and $P_{{\rm m11},\Delta k}\equiv P_{{\rm m11}}(\Delta k)$, etc., and defined 
\begin{align}
 a_{1} &=\left( b_{2,13}^{{\rm sq}} + b_{2,23}^{{\rm sq}}
   + 4 b_1 {\mathfrak{F}}_{2,{\rm eff}}^{{\rm sq}} \right) P_{{\rm m11}}P_{{\rm m11},\Delta k}\,, \\ 
a_{2} &=\left( 2b_{1,\Delta k}f_1 - {\mathfrak{G}}_{2,{\rm eff}}^{\rm sq} \right) P_{{\rm m11}}P_{{\rm m11},\Delta k} \,.
\end{align}
The bar indicates the ratio $\bar B_{02}^{\rm sq} =  {\cal P}_{1,\Delta k}^{-1} B_{02}^{\rm sq}$ etc., where
${\cal P}_{1,\Delta k} = f_{1,\Delta k}/ b_{1,\Delta k}$ is the linear observable in the $\Delta k$ mode, 
which is assumed to be in the quasilinear regime. 
As promised above, ${\cal P}_{1,\Delta k}$ is thus to be understood as an 
operator acting on given functions.
By contrast, we do not make use of the operator ${\cal P}_1 \equiv {\cal P}_1(k)$ 
as the $k$-mode could be in the quasilinear regime where the operator ${\cal P}_1$ delivers possibly a poor approximation of the underlying physics.
We have defined  
$2{\mathfrak{F}}_{2,{\rm eff}}^{{\rm sq}} \equiv {\mathfrak{F}}_{2,13}^{\rm sq} + {\mathfrak{F}}_{2,23}^{\rm sq}$
and
$2{\mathfrak{G}}_{2,{\rm eff}}^{{\rm sq}} \equiv {\mathfrak{G}}_{2,13}^{\rm sq} + {\mathfrak{G}}_{2,23}^{\rm sq}$
 which are free of infrared divergences even in the vicinity of $\Delta k \to 0$. 
We note that in deriving the above expressions, we have assumed 
that ${\mathfrak{F}}_{2,12} = {\mathfrak{F}}_2(\fett{k},-\fett{k}) =0$, a relation which is trivial 
to see in an EdS universe but generally holds also in DE models, as we shall prove 
in Appendix~\ref{app:squezzed}.

The galaxy bispectrum coefficients contain mostly too cluttered information 
about unknowns, and this is why we investigate in the following more sources 
of potential observables. Nevertheless, some of the above coefficients will 
become essential when determining our observables.

%%%%%%%%%%%%%%%%%%%%%%%%%%%%%%%%%%%%%%%%%%%%%%%%%%%%%%%
\section{Lensing and Lensing-Galaxy cross-spectra }\label{sec:Lens-GaBisp} 

Weak lensing, together with cross-correlations, provides another important tool 
in our analysis to gain further knowledge of quasilinear structure formation.
To discuss weak lensing we make use of the scalar line element
${\rm d}s^{2}= -\left(1+2\Psi\right) \dd t^2 +a^2\left(1+2\Phi\right) {\rm d}{\fett x}^2$
up to second order. We neglect
vector and tensor modes as we are usually interested in 
DE modifications of the scalar type. Secondary vector and tensor modes, 
even present in standard $\Lambda$CDM
cosmologies (see e.g.\ \cite{Rampf:2014mga}), are ignored as well, as their 
impact should be vanishingly small on the scales we consider.

Dark energy models usually modify the source term in the Poisson 
equation~(\ref{eq:modPois}), and on top of that, modifications in the 
gravitational slip are expected as well, the latter defined by
\begin{equation}
 \eta = -\frac{\Phi}{\Psi} \,.
\end{equation}
In $\Lambda$CDM we have $\eta \to 1$, however, only to first order in perturbation 
theory, and when ignoring massive neutrinos and the effects of baryons.
As regards to the impact of baryons, as we do limit our analysis to sufficiently 
large scales where baryonic effects should be small (see our assumption (b)),  we nevertheless expect that for weakly nonlinear scales, a mild baryonic impact could be incorporated in our framework. 
In any case, as mentioned above, we leave $\eta$ as a free function and do not invoke 
any specific parametrization.

Gravitational lensing is unaffected by redshift-space distortions or 
the (unknown) bias,
and is instead only sensitive to the total matter perturbation,
\begin{equation}
 k^{2}\Phi_{{\rm lens}}=k^2 \left( \Psi - \Phi \right) = -\frac 3 2 \Sigma\,\Omega_{{\rm m}}\delta_{\rm m} \,,
\end{equation}
where we have defined the modified lensing function $\Sigma=Y\left(1+\eta\right)$, with $Y \to 1$, however, only in the ``simplistic'' $\Lambda$CDM model (see above) which we do not assume.
What we truly observe in a measurement of gravitational lensing is the projection 
of the three-dimensional power spectrum and the bispectrum on a two-dimensional sphere integrated along the line 
of sight. The integral involves a window function that depends on the survey 
specification and geometry of the background space-time. Assuming a perfect 
knowledge of the window function one can differentiate the integral relation between 
the 3D and the 2D spectra and therefore link the unprojected 3D bispectrum to the 
actual observations.

%%%%%%%%%%%%%%%%%%%%%%%%%%%%%%%%%%%%%%%%%%%%%%%%%%%%%%%
\subsection{Lensing bispectrum}

We define the lensing bispectrum as 
\begin{align}
 &\Omega_{{\rm m}}^{3}\Big\langle\Sigma(k_{1})\delta_{\rm m}(\fett{k}_{1})\,\Sigma(k_{2})\delta_{\rm m}(\fett{k}_{2})\,\Sigma(k_{3})\delta_{\rm m}(\fett{k}_{3})\Big\rangle_{{\rm c}}\equiv  \nonumber \\
    &\quad(2\pi)^{3}\delta_{{\rm D}}^{(3)}\left(\fett{k}_{123}\right)B_{{\rm lens}}(k_{1},k_{2},k_{3}) \,.
\end{align}
Since the lensing signal is not sensitive to redshift-space distortions, 
the lensing bispectrum will not be affected by any projection effects.
We obtain at the leading order for the equilateral configuration 
\begin{align}
 B_{\rm lens}^{\rm eq} &=  \Omega_{\rm m}^3 \Sigma^3 \left( 6{\mathfrak{F}}_2^{\rm eq} P^2_{\rm m 11} 
   + B_{\rm m111}^{\rm eq} \right)\,,
\intertext{and for the squeezed configuration} 
 B_{\rm lens}^{\rm sq} &=  \Omega_{\rm m}^3 \Sigma^2 \Sigma_{\Delta k} \left( 4 {\mathfrak{F}}_{2, \rm eff}^{\rm sq} P_{\rm m11} P_{\rm m11, \Delta k} + B_{\rm m111}^{\rm sq}  \right) \,.
\end{align}

%%%%%%%%%%%%%%%%%%%%%%%%%%%%%%%%%%%%%%%%%%%%%%%%%%%%%%%
\subsection{Lensing-galaxy cross bispectra}

We also consider two types of cross-correlations between galaxy
and lensing signal, the first is the galaxy-galaxy-lensing bispectrum,
defined by
\begin{align}
&\Omega_{{\rm m}}\Big\langle\,\delta_{{\rm g}}^{{\rm s}}(\fett{k}_{1})\,\delta_{{\rm g}}^{{\rm s}}(\fett{k}_{2})\,\Sigma(k_{3})\delta(\fett{k}_{3})\,\Big\rangle_{{\rm c}}\equiv  \nonumber \\
&\quad (2\pi)^{3}\delta_{{\rm D}}^{(3)}(\fett{k}_{123})B^{{\rm ggl}}(\fett{k}_{1},\fett{k}_{2},\fett{k}_{3})\,,
\end{align}
which is in the equilateral configuration 
\begin{equation}
  B^{\rm ggl, eq} =  \Omega_{\rm m} \Sigma  P_{\rm m11}^2 \sum_{i}  B^{\rm ggl, eq}_{i} \mu_1^i \,,
\end{equation}
with nonvanishing coefficients  $B^{\rm ggl, eq}_{0}$, $B^{\rm ggl, eq}_{2}$, $B^{\rm ggl, eq}_{4}$, and $B^{\rm ggl, eq}_{6}$.  All coefficients are reported in 
Appendix~\ref{app:bispec}, in the following we only need the last one, i.e.,  
\begin{align}
 B^{\rm ggl, eq}_6 &= - \frac{59}{128}  f_1^3 \,.
\end{align}
We have also derived the squeezed limit of that cross-bispectra, together with
the other cross-bispectrum, the lensing-lensing-galaxy bispectrum, defined by
\begin{align}
&\Omega_{{\rm m}}^{2}\Big\langle\,\Sigma(k_{1})\delta(\fett{k}_1)\,\Sigma(k_2)\delta(\fett{k}_{2})\,\delta_{{\rm g}}^{{\rm s}}(\fett{k}_{3})\,\Big\rangle_{{\rm c}}\equiv  \nonumber \\
&\quad (2\pi)^{3}\delta_{{\rm D}}^{(3)}(\fett{k}_{123})B^{{\rm llg}}(\fett{k}_{1},\fett{k}_{2},\fett{k}_{3}) \,,
\end{align}
and we report all coefficients in Appendix~\ref{app:bispec}.
We note that in deriving the equilateral coefficients for the cross-correlators $B^{\rm ggl}$ and $B^{\rm llg}$, we have integrated out the azimuthal angular dependence $\phi$ 
as explained around Eq.\,(\ref{eq:cosinesEqui}). 

A comment on stochastic biasing models (see e.g.\,\cite{Dekel:1998eq}) is in order. 
In that class of phenomenological models, one introduces correlation coefficients, usually dubbed $r$, that parametrize the disknowledge of the underlying deterministic
formation process of biased tracers \cite{Matsubara:2011ck}. 
Since we do not assume any simplified bias model, 
our nonlocal bias model incorporates the stochasticity between matter and galaxy fields, and thus, we do not need to introduce these correlation coefficients for our cross-correlators. See Sec.\,II E in Ref.\,\cite{Matsubara:2011ck} for a highly related discussion.

%%%%%%%%%%%%%%%%%%%%%%%%%%%%%%%%%%%%%%%%%%%%%%%%%%%%%%%
\section{Observables} \label{sec:Observables}

The cosine-independent coefficients of the various sorts of bispectra 
are not directly observable since they are proportional to the model-dependent matter power spectrum and to the unknown normalization of the density fluctuation amplitude. However, taking ratios of these coefficients, these unknowns drop out.
 Taking the time derivative of a coefficient by subsequent division by a coefficient is another useful operation, since unknowns disappear.
 Thus, this methodology provides access to a wealth of cosmological information in a model-independent way.

In the following we briefly summarize the findings of linear observables that can be obtained from the galaxy and lensing power spectrum,
then we extend the set of observables into the quasilinear regime, by the use of the above bispectrum coefficients.

%%%%%
\subsection{Linear observables} \label{sec:linearO}

This section summarizes the findings from the literature \cite{Dodelson:2003ft,Zhang:2007nk,Percival:2008sh,Stebbins:2012vw,Cole:1993kh,Hatton:1997xs}, and we follow in particular the procedure of Ref.~\cite{Amendola:2012ky}. There it has been shown that 
taking the ratio of the power spectrum coefficients $P_2$ and $2P_4$ (see Eq.\,(\ref{powercoeff})), one gets $b_1/f_1$, whereas taking the time derivative of $P_4$ divided by $P_4$ gives $f_{1}+f_{1}'/f_{1}$.
Another important linear observable is $\Omega_{\rm m}\Sigma/f_1$, which is obtained by taking the ratio of the lensing power spectrum and $P_4$. 
In summary, the linear observables are \cite{Amendola:2012ky}
\begin{align} \label{linearobs}
 \begin{aligned}
{\cal P}_1 &= f_1 / b_1 \,,\qquad\qquad 
   {\cal P}_2 = \Omega_{{\rm m0}} \Sigma / f_1 \,,  \\
{\cal P}_3 & =f_1+f_1'/f_1 \,.
  \end{aligned} 
\end{align}
Interestingly, we obtain these (and many more) observables also from the bispectrum coefficients, with the important difference, that the bispectrum coefficients should hold on a wider range of scales, simply because they are obtained by using a better approximation in perturbation theory. 

The above linear observables are well known in the literature, and we note that
${\cal P}_1$ is often denoted with $\beta$ \cite{Cole:1993kh,Hatton:1997xs}, whereas ${\cal P}_2$ is sometimes called $E_G$ \cite{Zhang:2007nk}.

%%%%%
\subsection{Quasilinear observables}

It is straightforward to confirm from our bispectrum coefficients the findings of ${\cal P}_1$--${\cal P}_3$, but now obtained from a wider range of cosmological scales,
\begin{align}
 {\cal B}_1 & = \frac{B_{40}^{\rm sq} - \bar B_{42}^{\rm sq}}{B_{00}^{\rm sq} - \bar B_{02}^{\rm sq}} = \frac{f_1^2}{b_1^2}  \,,  \label{obsB1}\\
  {\cal B}_2 &=  - \frac{87 {\cal E}^2 \Omega_{\rm m} \Sigma B_{6}^{\rm ggl, eq}}{472 (1+z)^3 B_8^{\rm eq}} =  \frac{\Omega_{\rm m0} \Sigma}{f_1} \,,  \\
 {\cal B}_{3} &= \frac 1 4 \frac{\left( B_8^{\rm eq} P_{\rm m1}^2 \right)'}{B_8^{\rm eq} P_{\rm m1}^2 } = f_1 + \frac{f_1'}{f_1}  \label{timederivative}  \,.
\end{align}
%Evidently, these observables are constructed from weakly nonlinear quantities but agree with the linear observables. 
The equivalence of these observables with ${\cal P}_1$--${\cal P}_3$ can be used to establish a consistency relation in various ways. For example if the actual measurements from both
the linear and quasilinear regime yield inconsistent results, there could be some unresolved systematic in the theory or analysis.

Note that the above observables are independent of the unknown intrinsic bispectrum contribution $B_{\rm m111}$ (or $Q_{\rm m111}$). 
In fact, since we do not want to specify the DE model, second-order perturbations
arising from the nonlinear matter evolution are indistinguishable from PNG modifications, as it is also evident from the following two observables,
\begin{align}
 {\cal B}_{4} &= - \frac{29B_{\rm lens}^{\rm eq}}{2048B_8^{\rm eq} P_{\rm m11}^2 {\cal B}_3^3}   =  f_1^{-1} {\mathfrak{F}}_2^{\rm eq} + \frac{Q_{\rm m111}^{\rm eq}}{6f_1} \,, \\
 {\cal B}_{5} &= \frac{29B_6^{\rm eq}}{944B_8^{\rm eq}} -\frac{1}{16}  = f_1^{-2}{\mathfrak{G}}_2^{\rm eq} - \frac{Q_{\rm m111}^{\rm eq}}{6f_1}  \,.  
\intertext{However, summing up these two observables, we obtain another important 
observable that is independent of $Q_{\rm m111}$,}
  {\cal B}_{6} &=  {\cal B}_{4} +  {\cal B}_{5} = f_1^{-1} {\mathfrak{F}}_2^{\rm eq} + f_1^{-2}{\mathfrak{G}}_2^{\rm eq} \,. \label{obsB6-B7}
\end{align}
This observable will become crucial in the following section when we establish a nonlinear model-independent constraint.

What knowledge can be gained about the nonlocal bias coefficients? We find model-independent constraints for the following bias ratios,
\begin{align}
  {\cal B}_7 &= \frac{b_2^{\rm eq}}{b_1^2} \,, \qquad {\cal B}_8  = \frac{b_{2,12}^{\rm sq}}{b_1^2} \,, \label{biasratio}
\end{align}
which we shall derive in Appendix~\ref{app:moreObs}, where we also provide even more observables. What is missing is a similar uncluttered observable involving $b_{2,13}^{\rm sq}$ or $b_{2,23}^{\rm sq}$, which, however, we have been unable to find. \\[0.3cm]

%\begin{widetext}
%%%%%%%%%%%%%%%%%%%%%%%%%%%%%%%%%%%%%%%%%%%%%%%%%%%%%%%%
\section{Model-independent constraints}\label{sec:consistency} 

\subsection{Linear regime}
Observe that the PDE for the linear matter density, Eq.\,(\ref{evoLinearDelta}), can be rewritten in terms of a PDE for $f_1$,
\begin{equation}
  f_{1}'+f_{1}^{2}+f_{1}\left(2+\frac{{\cal E}'}{\cal E}\right) =\frac{3}{2}\Omega_{{\rm m}}Y  \label{delta1evo} \,,
\end{equation}
where we remind the reader that ${\cal E} = H/H_0$, and we have
\begin{equation}
\Omega_{\rm m} =\Omega_{\rm m0} \frac{(1+z)^3} {{\cal E}^2}\,.
\end{equation}
In Ref.\,\cite{Amendola:2012ky} it has been shown, using the set of linear observables~(\ref{linearobs}), that the above equation turns into a relation for
the anisotropic stress $\eta$,
\begin{equation} 
  \frac{3 {\cal P}_2 (1+z)^3}{2 {\cal E}^2 \left( {\cal P}_3 + 2 + {\cal E}'/{\cal E}\right)} -1 = \eta \,. \label{linearconsistency}
\end{equation}
This relation implies a model-independent constraint of $\eta$ in terms of linear observables, a powerful result that can be used e.g.\ to rule out entire classes of DE models.

\subsection{Quasilinear regime}\label{sec:nonlinearconstraint}

Here we seek a similar relation as above, now obtained, however, from our novel
quasilinear observables. To achieve this,
we use the linear result $\theta_{\rm m1} = - f_1 \delta_{\rm m1}$ and the fully general
ansatz (cf.\ Eqs.\,(\ref{pertAnsatz}); here suppressing the integrals and 
Dirac deltas because of notational simplicity)
\begin{equation}
  \delta_{\rm m2} = {\mathfrak{F}}_2\, \delta_{\rm m1} \delta_{\rm m1}\,, 
 \qquad   \theta_{\rm m2} = {\mathfrak{G}}_2\, \delta_{\rm m1} \delta_{\rm m1}
\end{equation}
in Eqs.\,(\ref{eq:conti})--(\ref{eq:mom}) together with the modified Poisson equation~(\ref{eq:modPois}). Taking the divergence of Eq.\,(\ref{eq:mom}),
expanding Eqs.\,(\ref{eq:conti})--(\ref{eq:mom}) in perturbation theory 
and Fourier transforming the
resulting expressions, these equations become, respectively, at second order 
\begin{widetext}
\begin{align}
 \left\{ {\mathfrak{F}}_2' + {\mathfrak{F}}_2 \left[ f_1(k_1) + f_1(k_2) \right] \right\} \delta_{\rm m1} \delta_{\rm m1} &=  \left\{ \frac 1 2 \frac{\fett{k}_1 \cdot \fett{k}_2}{k_1 k_2} \left[ \frac{f_1(k_1)k_2}{k_1} + \frac{f_1(k_2)k_1}{k_2} \right]
  - {\mathfrak{G}}_2 + \frac 1 2 f_1(k_1)+ \frac 1 2 f_1(k_2) \right\} \delta_{\rm m1} \delta_{\rm m1} \,,  \label{contiF2}  \\
 \left\{ {\mathfrak{G}}_2' + {\mathfrak{G}}_2 \left[ f_1(k_1) + f_1(k_2) \right] \right\} \delta_{\rm m1}\hspace{0.05cm} \delta_{\rm m1} &= \Bigg\{ - \left( 2+ \frac{{\cal E}'}{\cal E}\right) {\mathfrak{G}}_2 - f_1(k_1) f_1(k_2) \left( \frac{\fett{k}_1 \cdot \fett{k}_2}{k_1 k_2}\right)^2 \nonumber \\
  &\qquad\hspace{1.9cm}- \frac 1 2 f_1(k_1) f_1(k_2) \frac{\fett{k}_1 \cdot \fett{k}_2}{k_1 k_2} \left[ \frac{k_1}{k_2} + \frac{k_2}{k_1} \right]
  - \frac 3 2 \Omega_{\rm m} Y {\mathfrak{F}}_2 \Bigg\} \delta_{\rm m1} \delta_{\rm m1} \,. \label{eulerG2}
\end{align}
These relations must also hold for specific configurations and without loop 
integrals (see App.\,\ref{app:nointegral} for a rigorous proof).
For example, in the equilateral case we get the two relations
\begin{align}
  {{\mathfrak{F}}_2^{\rm eq}}' + 2 f_1 {\mathfrak{F}}_2^{\rm eq} &= \frac{f_1}{2} - {\mathfrak{G}}_2^{\rm eq} \,, \label{contiF2eq}\\
  {{\mathfrak{G}}_2^{\rm eq}}' + 2 f_1 {\mathfrak{G}}_2^{\rm eq} &=  - \left( 2 + \frac{{\cal E}'}{\cal E} \right) {\mathfrak{G}}_2^{\rm eq} 
           + \frac{f_1^2}{4} - \frac 3 2  \Omega_{\rm m } Y {\mathfrak{F}}_2^{\rm eq}\,. \label{eulerG2eq}
\end{align}
\end{widetext}
Now, making use of these equations, and the quantity 
$f_1^2 {\cal B}_6 = f_1 {\mathfrak{F}}_2^{\rm eq} + {\mathfrak{G}}_2^{\rm eq}$ with its time derivative, 
${{\mathfrak{G}}_2^{\rm eq}}' = \left(f_1^2 {\cal B}_6 \right)' - f_1' {\mathfrak{F}}_2^{\rm eq} - f_1 {{\mathfrak{F}}_2^{\rm eq}}'$,
we obtain a model-independent realization of $f_1$.
For this we first use Eq.\,(\ref{contiF2eq}) to get an expression for ${{\mathfrak{F}}_2^{\rm eq}}'$, and then plug this into the expression for ${{\mathfrak{G}}_2^{\rm eq}}'$ in terms of ${\cal B}_6$. We get
\begin{equation}
  {{\mathfrak{G}}_2^{\rm eq}}' =  \left(f_1^2 {\cal B}_6 \right)' - f_1' {\mathfrak{F}}_2^{\rm eq}
     - \frac{f_1^2}{2} + f_1^3 {\cal B}_6 + f_1^2 {\mathfrak{F}}_2^{\rm eq}  \,.
\end{equation}
Plugging this in~(\ref{eulerG2eq}) we finally get after a little algebra
\begin{equation} \label{eq:consistency}
   \frac{3}{4{\cal B}_6} - \frac{{\cal B}_6'}{{\cal B}_6}  - 2 {\cal B}_3 -  \left( 2 + \frac{{\cal E}'}{{\cal E}} \right) = f_1 \,, 
\end{equation}
where we have used Eqs.\,(\ref{timederivative}) and~(\ref{delta1evo}). This is our main result. We stress that the lhs is obtained from model-independent observables, and thus this equation delivers a model-independent measurement of $f_1$. Furthermore we note that Eq.\,(\ref{eq:consistency}) is independent of the modified Poisson source function $Y$, as the latter drops out during the derivation of~\eqref{eq:consistency}.

Having obtained $f_1$, we get from the quasilinear observables~(\ref{obsB1}) and~(\ref{biasratio}) the bias parameters $b_1$, $b_2^{\rm eq}$ and $b_{2,12}^{\rm sq}$ as well as the quantity $\Omega_{\rm m0} \Sigma$. 
%Finally, once $f_1$ and its time derivative has been reconstructed, one obtains, by evaluating the lhs of Eq.\,(\ref{delta1evo}), also the quantity $\Omega_{\rm m0} Y$.
%
If we furthermore use the linear relation~(\ref{linearconsistency}) that gives $\eta$, we also get $\Omega_{\rm m0} Y$, by virtue of $\Sigma = Y(1+\eta)$, i.e., 
\begin{equation}
\Omega_{\rm m0} Y=\frac {2 f_1 {\cal B}_2 {\cal E}^2 ({\cal P}_3+2+{\cal E'}/{\cal E})}{3{\cal P}_2 (1+z)^3} \,.
\end{equation}
These are our final results, and we remind the reader that they are valid under assumptions (a)--(e), see Sec.\,\ref{sec:setup}.

\newpage
%%%%%%%%%%%%%%%%%%%%%%%%%%%%%%%%%%%%%%%%%%%%%%%%%%%%%%%%
\section{Challenges of perturbation theory}\label{sec:valid} 

The linear and quasilinear observables, together with the constraint equations 
have been obtained  within the framework of SPT, the latter
being based on a single-stream fluid description which breaks down when particle
trajectories begin to intersect. At that instant, the fluid enters the
multi-stream regime and velocities become multi-valued, which evidently excites
higher-order kinetic moments of the Vlasov hierarchy ---
 such as the velocity dispersion
tensor (see e.g.\ \cite{Bernardeau:2001qr}). 
Also, even if the fluid was initially curlfree, vorticities are generated in the
multi-stream regime.
Both the presence of a nonvanishing velocity dispersion and vorticity could 
restrict the validity of parts of the above calculation to sufficiently large scales,
although it is expected that both effects hamper the analysis  deep in the nonlinear regime the most.

Let us first elucidate the consequences of the presence of vorticity generation in 
the multi-stream regime, that we have neglected in the present paper. 
Nonvanishing vorticity implies that the velocity cannot be described by just its divergence $\theta_{\rm m}$. As a result, the velocity power spectrum is a superposition
of two power spectra, one for its divergence and the other for its noncurlfree part.
Since our observables make use only of the divergence part of the total
velocity power spectrum, one could question whether the estimators could be 
biased in the presence of vorticity. The effect of vorticity on the total velocity 
power spectrum has been investigated by a suite of cosmological simulations in 
Ref.\,\cite{Pueblas:2008uv}. 
There it has been shown that at late times ($z=0$), 
the amplitude of the vorticity power spectrum
is by a factor of about 250 smaller compared to the one from the divergence part for scales larger than $0.4h/$Mpc (see their Fig.\,3;  notice, however, the residual dependence on the mass resolution for the extraction of the vorticity power spectrum).
Thus, for sufficiently large scales only little power gets transferred to the curl part of the velocity, and vorticity can be safely neglected.  

The next issue we discuss
 is the effect of velocity dispersion in the onset of multi-streaming. 
Incorporating velocity dispersion is generally a small-scale problem that must be modelled deep in the multi-stream regime, but in several numerical studies it has been shown that redshift matter polyspectra could be affected on mildly nonlinear or even linear scales (e.g., \cite{Scoccimarro:1999ed,Pueblas:2008uv}). 
Also, not only the matter density but also the matter velocity divergence 
receives corrections induced through velocity dispersion, 
and the respective feedbacks have been assessed in Ref.\,\cite{Pueblas:2008uv}.
There the authors report that at late times 1\% corrections arise 
on the velocity divergence matter power spectrum at scales smaller than $0.1h/$Mpc. 
Estimating the corrections induced through velocity dispersion 
is, however, a difficult task, and in Ref.\,\cite{Pueblas:2008uv} the authors 
have applied only a linearization-based estimate of the impact of velocity 
dispersion.
Furthermore, it remains unclear whether the measured velocity
dispersion in these simulations is physical or remnants of finite 
resolution effects \cite{Cusin:2016zvu}. 

Accurate theoretical modelling of the effect of velocity dispersion on redshift polyspectra 
is still an open problem, although considerable progress has been made in the past years. Advanced models make use for example of resummation schemes that resum the infinite SPT series in Lagrangian space \cite{Matsubara:2007wj,Rampf:2012xb}, or for example the distribution function approach \cite{Seljak:2011tx,Vlah:2012ni} that uses an extended version of perturbation theory to capture velocity dispersion effects more accurately than SPT.
The simplest models, by contrast, are motivated by phenomenological considerations and
modify the redshift galaxy power- and bispectrum by hand (e.g., \cite{Peacock:1993xg,Taruya:2010mx,Hashimoto:2017klo}).
All the models so far in the literature essentially introduce a suppression factor in the power- and bispectrum. These suppression factors have in common that all of them affect mostly the overall shape of the polyspectra, whereas leaving other features such as the ``wiggle information'' (i.e., baryonic acoustic oscillations) almost unaltered (see e.g., Fig.\,2 of Ref.\,\cite{Rampf:2012xb}). 
Thus, when restricting to sufficiently large scales, we expect that our observables are mostly unaffected by velocity dispersion, since the shape information cancels out when taking the ratios of different $\mu$ coefficients of the bispectra.

 A significant nonzero velocity dispersion, however, 
would alter the momentum conservation of matter, i.e., the velocity dispersion tensor 
would explicitly appear in Eq.\,(\ref{eq:mom}) and consequently also in~\eqref{eq:consistency}. 
Fortunately, our observables are measured from galaxy samples and not from the dark matter distribution directly, 
and it is known that velocity dispersion effects are generally smaller 
for galaxies than for matter, especially if a sample of central galaxies can be selected \cite{Vlah:2012ni,Okumura:2012xh}.
Nonetheless we believe that the assumption of small velocity dispersion in the present paper is the most stringent one, i.e., the one which limits the validity of the present approach to sufficiently large scales. We thus consider further theoretical investigations in this direction as an important task, but beyond the scope of the present study.

%%%%%%%%%%%%%%%%%%%%%%%%%%%%%%%%%%%%%%%%%%%%%%%%%%%%%%%%
\section{Conclusions}\label{sec:concl}

We have shown that, without imposing any DE parametrization, cosmological observations 
can measure only 
(1) $\Omega_{\rm k0}$ and ${\cal E} = H/H_0$ at the background level; 
(2) the combinations ${\cal P}_{1} =f_{1}/b_{1}$, ${\cal P}_{2}=\Omega_{{\rm m0}}\Sigma/f_{1}$
    and ${\cal P}_{3} =f_{1}+f_{1}'/f_{1}$ at the linear level; and 
(3) the novel observables ${\cal B}_1 - {\cal B}_{10}$ and ${\cal C}_1-{\cal C}_4$ that 
are applicable in the quasilinear regime. (A concise list of these nonlinear 
observables is given in Appendix~\ref{app:moreObs}.)
The observables ${\cal P}_{1} - {\cal P}_{3}$ and ${\cal B}_1 - {\cal B}_{3}$ are formally 
identical, with the difference that the former are obtained from a strictly linear analysis,
whereas the latter includes the leading nonlinearities.
However, the quasilinear observables
also apply to the linear scales and thus, our quasilinear observables can probe a 
larger range of scales than could be done with a strictly linear analysis.
Furthermore, applying both the linear and quasilinear observables 
to linear scales only,
the respective measurements of the observables must deliver identical results. 
From this one could perform consistency tests that rule 
out entire classes of DE models.

Many unknowns remain unknowns, especially $\Omega_{\rm m0}$, the DE density 
parameter $\Omega_{\rm x}$, and we are left with a degeneracy between non-Gaussianities 
in the initial conditions (arising from PNG) and non-Gaussianities 
from the matter evolution.
From our nonlinear observables, however, we can derive a model-independent constraint 
equation given in Eq.\,(\ref{eq:consistency}). This relation should hold for a wide range 
of DE models, and, if verified by cosmological observations, can be used to obtain a
 model-independent measure of $f_1$. That in turn, in combination with our
observables, enables us to reconstruct the bias parameters $b_1$ and $b_2$, and
the quantities $\Omega_{\rm m0}Y$ and $\Omega_{\rm m0}\Sigma$. Lastly, having $f_1$
one gets the normalization-dependent quantity 
$R= D f_1 \sigma_8 \delta_{\rm m0}$ \cite{Amendola:2012ky}, from which one 
gets $\sigma_8^2 P_{\rm m1}$ as well.

All the practical limitations of a real measurement, that we neglect here, are 
of course the most challenging problem to handle 
(see e.g., \cite{Levi:2013gra,Amendola:2016saw,Bull:2015stt}).
This paper, thus, should be understood as a potential starting point for a long journey
with many hurdles ahead, with the final goal to reconstruct or reject 
entire classes of DE cosmologies.

\section*{ACKNOWLEDGEMENTS}

C.R.\ thanks V.\ Desjacques for useful discussions.
The work of C.R.\ and L.A.\ is supported by the DFG through the Transregional Research Center TRR33 ``The Dark Universe.''
C.R.\ acknowledges the support of the individual Grant No.\ RA 2523/2-1 from the DFG.
E.V.\ thanks the INFN-INDARK initiative Grant No.\ IS PD51 for financial
support.  

%\newpage

%%%%%%%%%%%%%%%%%%%%%%%%%%%%%%%%%%%%%%%%%%%%%%%%%%%%%%%

\appendix
%dummy comment inserted by tex2lyx to ensure that this paragraph is not empty

%%%%%%%%%%%%%%%%%%%%
\section{All bispectrum coefficients}\label{app:bispec}

In the main text we have mentioned only bispectrum coefficients that are relevant in deriving our main results, while skipping other coefficients.
Here we give a complete list of bispectrum coefficients.

The galaxy bispectrum in the equilateral configuration reads
\begin{equation} 
B_{{\rm g}}^{{\rm eq}}= P_{{\rm m11}}^{2} \sum_{i}B_{i}^{{\rm eq}}\,\mu_{1}^{i}\,,
\end{equation}
with the nonvanishing coefficients
\begin{align}
 B_0^{{\rm eq}} &= \frac{27}{128} f_1^2 b_2^{\rm eq} + \frac{3}{4}  f_1 b_1^3 + 6 b_1^3 {\mathfrak{F}}_2^{\rm eq} + 3 b_1^2 b_2^{\rm eq}+ \frac{27}{64}  f_1^2 b_1^2  \nonumber \\ 
 &\!\!\!\!+ 3 f_1 b_1^2  {\mathfrak{F}}_2^{\rm eq}  -\frac{3}{2}b_1^2{\mathfrak{G}}_2^{\rm eq}  + \frac{3}{2} f_1 b_1 b_2^{\rm eq}+\frac{27}{64} f_1^2 b_1 {\mathfrak{F}}_2^{\rm eq} \nonumber \\ 
 &\!\!\!\!-\frac{27}{32} f_1 b_1 {\mathfrak{G}}_2^{\rm eq} 
  + Q_{\rm m111}^{\rm eq} \!\!\left( b_1^3 + \frac 3 4 b_1^2 f_1 + \frac{27b_1f_1^2}{128} \right)\! , \\ 
B_2^{{\rm eq}} &= \frac{3}{4} f_1 b_1^3  + \frac{9}{32} f_1^2 b_1^2 +  3 f_1 b_1^2  {\mathfrak{F}}_2^{\rm eq} - \! \frac 3 2 b_1^2 {\mathfrak{G}}_2^{\rm eq} + \! \frac{3}{2} f_1 b_1  b_2^{\rm eq} \nonumber \\ 
   &\!\!\!\!+ \frac{9}{32} f_1^2 b_1 {\mathfrak{F}}_2^{\rm eq} -\frac{9}{16} f_1 b_1 {\mathfrak{G}}_2^{\rm eq}  + \frac{9}{64} f_1^2 b_2^{\rm eq} - \frac{135}{1024} f_1^4 \nonumber  \\ 
   &\!\!\!\!-   \frac{81}{64} f_1^2{\mathfrak{G}}_2^{\rm eq} + \frac{3f_1}{128} Q_{\rm m111}^{\rm eq} \left( 32 b_1^2 + 6 b_1 f_1 + 9 f_1^2 \right) , \\ 
 B_4^{{\rm eq}} &=  \frac{27}{128} f_1^2 b_2^{\rm eq} +\frac{351}{1024} f_1^4 + \frac{27}{64} f_1^2 b_1 {\mathfrak{F}}_2^{\rm eq} -\frac{27}{32} f_1 b_1 {\mathfrak{G}}_2^{\rm eq} \nonumber  \\ 
   &\!\!\!\!+ \frac{27}{64} f_1^2 b_1^2 +\frac{117}{32}f_1^2{\mathfrak{G}}_2^{\rm eq}  
 + \frac{3f_1^2}{128} Q_{\rm m111}^{\rm eq} \left( 9 b_1 - 26 f_1 \right) ,\\  
B_6^{{\rm eq}} &=-\frac{177}{1024}  f_1^2  \left( f_1^2 + 16 {\mathfrak{G}}_2^{\rm eq} -\frac{8}{3}Q_{\rm m111}^{\rm eq} f_1 \right) \,, \\
B_8^{{\rm eq}} &= -\frac{87}{1024}f_1^4 \,,
\end{align}
where
\begin{equation}
  Q_{\rm m111}^{\rm eq} \equiv B_{\rm m111}^{\rm eq} /P_{\rm m11}^2 \,.
\end{equation}

For the squeezed galaxy bispectrum we have
\begin{align}
B_{{\rm g}}^{{\rm sq}}=\sum_{i,j}B_{ij}^{{\rm sq}}\mu_{1}^{i}\mu_{\Delta k}^{j}\,,
\end{align}
with the only nonvanishing coefficients 
\begin{align}
B_{00}^{\rm sq} & = 
   a_1  b_{1} b_{1,\Delta k} + b_{2,12}^{\rm sq} b_{1}^{2} P_{{\rm m11}}^{2}
   + B_{\rm m111}^{\rm sq} b_1^2 b_{1, \Delta k}  \,,   \\
\bar B_{02}^{\rm sq} & =  a_1  b_{1} b_{1,\Delta k} + B_{\rm m111}^{\rm sq} b_1^2 b_{1,\Delta k} \,, \\
B_{20}^{\rm sq} & =  a_1  f_1  b_{1, \Delta k} + a_2 b_1 b_{1, \Delta k}   +2 b_{2,12}^{\rm sq} b_{1} f_{1} P_{{\rm m11}}^{2} \nonumber \\
  &+ 2 B_{\rm m111}^{\rm sq} f_1 b_1 b_{1,\Delta k}   \,, \\
\bar B_{22}^{\rm sq} & =  a_1 f_1 b_{1, \Delta k} + a_2 b_1 b_{1, \Delta k}
  + 2 B_{\rm m111}^{\rm sq}  f_1 b_1 b_{1,\Delta k}   \,, \\
B_{40}^{\rm sq} &=  a_2 f_1 b_{1, \Delta k} +  b_{2,12}^{\rm sq} f_{1}^{2} P_{{\rm m11}}^{2} 
   + B_{\rm m111}^{\rm sq} f_1^2 b_{1,\Delta k}   \,, \\
 \bar B_{42}^{\rm sq} &=  a_2 f_1 b_{1, \Delta k} + B_{\rm m111}^{\rm sq} f_1^2 b_{1,\Delta k} \,, 
\end{align}
where 
\begin{align}
a_{1} & =\left(b_{2,13}^{{\rm sq}}+b_{2,23}^{{\rm sq}}+4b_{1}{\mathfrak{F}}_{2,{\rm eff}}^{{\rm sq}}\right)P_{{\rm m11}}P_{{\rm m11},\Delta k}\,,\\
a_{2} & =\left(4f_{2}{\mathfrak{F}}_{2,{\rm eff}}^{{\rm sq}}-2[f_{1}+f_{1,\Delta k}/2]+2b_{1,\Delta k}f_{1}\right) \nonumber \\
 &\qquad \hspace{3.3cm} \times P_{{\rm m11}}P_{{\rm m11},\Delta k}\,.
\end{align}
%%%

For the pure lensing bispectra, we get in the equilateral configuration 
\begin{align}
B_{\rm lens}^{\rm eq} &=   \Omega_{\rm m}^3 \Sigma^3 \left( 6{\mathfrak{F}}_2^{\rm eq} P^2_{\rm m 11} + B_{\rm m111}^{\rm eq} \right)\,,
\intertext{and in the squeezed configuration} 
B_{\rm lens}^{\rm sq} &=   \Omega_{\rm m}^3 \Sigma^2 \Sigma_{\Delta k} \left( 4 {\mathfrak{F}}_{2, \rm eff}^{\rm sq} P_{\rm m11} P_{\rm m11, \Delta k} + B_{\rm m111}^{\rm sq}  \right) ,
\end{align}
which, evidently, have no angular dependence.

%%%
Next is the cross-bispectrum 'galaxy-galaxy-lensing' which is in the equilateral configuration 
\begin{equation}
  B^{\rm ggl, eq} =  \Omega_{\rm m} \Sigma  P_{\rm m11}^2 \sum_{i}  B^{\rm ggl, eq}_{i} \mu_1^i \,,
\end{equation}
with 
\begin{align}
B^{\rm ggl, eq}_{0} &=  \frac 3 8 f_1 b_2^{\rm eq} + \frac 3 8 f_1 b_1^2 + 6 b_1^2 {\mathfrak{F}}_2^{\rm eq} + 2 b_1 b_2^{\rm eq} + \frac 3 2 f_1 b_1 {\mathfrak{F}}_2^{\rm eq}  \nonumber \\
   &\!\!\!\! - \frac 3 4 b_1 {\mathfrak{G}}_2^{\rm eq}  + \frac 1 8 Q_{\rm m111}^{\rm eq} \left( 8 b_1^2 + 3 b_1 f_1 \right)  \,, \\
 B^{\rm ggl, eq}_{2} &=  \frac 7 8 f_1 b_1^2 + \frac 7 8  f_1b_2^{\rm eq} - \frac{27}{128} f_1^3 + \frac 3 4 f_1^2 {\mathfrak{F}}_2^{\rm eq} + \frac{9}{16} f_1^2 b_1 \nonumber \\
 &\!\!\!\! + \frac 7 2 f_1 b_1 {\mathfrak{F}}_2^{\rm eq}- \frac 7 4 b_1 {\mathfrak{G}}_2^{\rm eq} - \frac 3 2 f_1 {\mathfrak{G}}_2^{\rm eq}     \nonumber \\
  &\!\!\!\!+ \frac 1 8 Q_{\rm m111}^{\rm eq} \left( 7 b_1 f_1 + 3 f_1^2 \right)  \,, \\
 B^{\rm ggl, eq}_{4} &=  \frac{1}{16} f_1^2 b_1 + \frac{39}{64} f_1^3 - \frac 1 4 f_1^2 {\mathfrak{F}}_2^{\rm eq} + \frac 1 2 f_1 {\mathfrak{G}}_2^{\rm eq}  \nonumber \\
  &\!\!\!\! - \frac 1 8 Q_{\rm m111}^{\rm eq} f_1^2  \,, \\ 
 B^{\rm ggl, eq}_{6} &=- \frac{59}{128}  f_1^3 \,,
\end{align}
and in the squeezed configuration 
\begin{equation}
B^{{\rm ggl,sq}}= \Omega_{\rm m} \Sigma_{\Delta k} \sum_{i}B_{i}^{{\rm ggl,sq}}\mu_{1}^{i}\,,
\end{equation}
with coefficients 
\begin{align}
B^{\rm ggl, sq}_0 &=  b_1 a_1  + B_{\rm m111}^{\rm sq} b_1^2  \,, \\
  B^{\rm ggl, sq}_2 &=  f_1 a_1 + b_1 a_2  + 2 B_{\rm m111}^{\rm sq} b_1 f_1  \,, \\
  B^{\rm ggl, sq}_4 &= f_1 a_2  + B_{\rm m111}^{\rm sq} f_1^2 \,, 
\end{align}
with $a_{1}$ and $a_{2}$ as above.

The second cross-bispectrum we consider is the lensing-lensing-galaxy bispectrum, defined by
\begin{align}
&\Omega_{{\rm m}}^{2}\Big\langle\,\Sigma(k_{1})\delta(\fett{k}_{1})\,\Sigma(k_{2})\delta(\fett{k}_{2})\,\delta_{{\rm g}}^{{\rm s}}(\fett{k}_{3})\,\Big\rangle_{{\rm c}}\equiv  \nonumber \\
&\quad (2\pi)^{3}\delta_{{\rm D}}^{(3)}(\fett{k}_{123})B^{{\rm llg}}(\fett{k}_{1},\fett{k}_{2},\fett{k}_{3})\,,
\end{align}
which is in the equilateral configuration 
\begin{equation}
   B^{\rm llg, eq} = \Omega^2_{\rm m} \Sigma^2  P_{\rm m11}^2 \sum_{i}  B^{\rm llg, eq}_{i} \mu_1^i  \,,
\end{equation}
with 
\begin{align}
  B^{\rm llg, eq}_{0} &=  b_2^{\rm eq} +\frac 3 8 f_1 b_1 + 6 b_1 {\mathfrak{F}}_2^{\rm eq} + \frac 3 2 f_1 {\mathfrak{F}}_2^{\rm eq} - \frac 3 4 {\mathfrak{G}}_2^{\rm eq} \nonumber \\
  &+ Q_{\rm m111}^{\rm eq} \left( b_1 + \frac 3 8 f_1 \right)  \,,  \\
 B^{\rm llg, eq}_{2} &=  \frac{3}{16} f_1^2 - \frac 1 8 f_1 b_1   - \frac 1 2 f_1 {\mathfrak{F}}_2^{\rm eq} + \frac 1 4 {\mathfrak{G}}_2^{\rm eq} - \frac 1 8 Q_{\rm m111}^{\rm eq} f_1  \,,  \\
 B^{\rm llg, eq}_{4} &= - \frac{5}{16} f_1^2 \,,
\end{align}
and in the squeezed configuration 
\begin{equation}
B^{{\rm llg,sq}}= \Omega_{\rm m}^2 \Sigma^2  \sum_{i}B_{i}^{{\rm llg,sq}}\mu_{\Delta k}^{i} \,,
\end{equation}
with
\begin{align}
B^{{\rm llg, sq}}_0 &=  4b_{1,\Delta k} {\mathfrak{F}}_{2,\rm eff}^{\rm sq} P_{\rm m11} P_{\rm m11, \Delta k}  + b_{2,12}^{\rm sq} P_{\rm m11}^2 \nonumber \\
  &+ B_{\rm m111}^{\rm sq} b_{1,\Delta k} \,, \\
 B^{{\rm llg, sq}}_2 &=  4 f_{1,\Delta k} {\mathfrak{F}}_{2,\rm eff}^{\rm sq} P_{\rm m11} P_{\rm m11, \Delta k} + B_{\rm m111}^{\rm sq} f_{1,\Delta k}  \,.
\end{align}

%%%%%%%%%%%%%%%%%%%%
\section{More nonlinear observables}\label{app:moreObs}

Here we report the full list of nonlinear observables including their derivations,
\begin{align}
   {\cal B}_1 & = \frac{B_{40}^{\rm sq} - \bar B_{42}^{\rm sq}}{B_{00}^{\rm sq} - \bar B_{02}^{\rm sq}} = \frac{f_1^2}{b_1^2}  \,,  \label{eq:firstObs} \\
  {\cal B}_2 &= - \frac{87 {\cal E}^2 \Omega_{\rm m} \Sigma B_{6}^{\rm ggl, eq}}{472 (1+z)^3 B_8^{\rm eq}} =  \frac{\Omega_{\rm m0} \Sigma}{f_1} \,,  \\
 {\cal B}_{3} &= \frac 1 4 \frac{\left( B_8^{\rm eq} P_{\rm m1}^2 \right)'}{B_8^{\rm eq} P_{\rm m1}^2 } = f_1 + \frac{f_1'}{f_1}  \,, \\
 {\cal B}_{4} &= - \frac{29B_{\rm lens}^{\rm eq}}{2048B_8^{\rm eq} P_{\rm m11}^2 {\cal B}_3^3}   =  f_1^{-1} {\mathfrak{F}}_2^{\rm eq} + \frac{Q_{\rm m111}^{\rm eq}}{6f_1} \,, \\
 {\cal B}_{5} &= \frac{29B_6^{\rm eq}}{944B_8^{\rm eq}} -\frac{1}{16}  = f_1^{-2}{\mathfrak{G}}_2^{\rm eq} - \frac{Q_{\rm m111}^{\rm eq}}{6f_1}  \,, 
\end{align}
\begin{align}
 {\cal B}_{6} &=  {\cal B}_{4} +  {\cal B}_{5} = f_1^{-1} {\mathfrak{F}}_2^{\rm eq} + f_1^{-2}{\mathfrak{G}}_2^{\rm eq} \,, \\
   {\cal B}_{7} &=  \frac{ 3 {\cal B}_1^{1/2} {\cal C}_{1}/16 - 3/ 8 - 6{\cal B}_{4}}{ 3/ 8 + 2 {\cal B}_1^{-1/2}} = \frac{b_2^{\rm eq}}{b_1^2} \,,  \\
{\cal B}_{8} &= {\cal B}_1 {\cal C}_{2} (1+ {\cal C}_{3}) = \frac{b_{2,12}^{\rm sq}}{b_1^2} \,, \\
   {\cal B}_{9} &= \frac{B_{\rm lens}^{\rm sq} {\cal B}_{1,\Delta k}^{-1/2} {\cal C}_{4,\Delta k}^{-1}}{{\cal B}_1 {\cal C}_4^2 B_{00}^{\rm sq} - B_0^{\rm llg, sq}} =
      b_1 \frac{ 4 {\mathfrak{F}}_{2,\rm eff}^{\rm sq} + Q_{\rm m111}^{\rm sq}}{b_{2,13}^{\rm sq} + b_{2,23}^{\rm sq}} \,, \\
{\cal B}_{10} &= \frac{ G }{\bar B_{22}^{\rm sq}-2 \bar B_{42}^{\rm sq} {\cal B}_1^{-1/2} - G {\cal B}_1^{1/2}} \nonumber \\
  &=   \frac{ b_{2,13}^{\rm sq} + b_{2,23}^{\rm sq}}{4 f_1 {\mathfrak{F}}_{2,\rm eff}^{\rm sq} -2 f_1 b_{1,\Delta k} + 4 {\mathfrak{G}}_{2,\rm eff}^{\rm sq} }  \,, 
\end{align}
and 
\begin{align}
  {\cal C}_1 &= - \frac{5 B_2^{\rm llg,eq} + 3 B_4^{\rm llg,eq} }{ B_4^{\rm llg,eq}}
    + 2{\cal B}_1^{-1/2} \nonumber \\
    & = -8 f_1^{-1} {\mathfrak{F}}_2^{\rm eq} + 4 f_1^{-2} {\mathfrak{G}}_2^{\rm eq} 
  -2 f_1^{-1} Q_{\rm m111}^{\rm eq} \,, \\
 {\cal C}_2 &= - \frac{177}{1024} \frac{B_{40}^{\rm sq} - \bar B_{42}^{\rm sq}}{B_6^{\rm eq} P_{\rm m1}^2} = \frac{b_{2,12}^{\rm sq}} {f_1^2 + 16 {\mathfrak{G}}_2^{\rm eq}  - 8 f_1 Q_{\rm m111}^{\rm eq}/3} \,, \\
  {\cal C}_3  &= \frac{87B_6^{\rm eq}}{177B_8^{\rm eq}} -1 = 16 f_1^{-2}  {\mathfrak{G}}_2^{\rm eq} - \frac 8 3 f_1^{-1}  Q_{\rm m111}^{\rm eq}  \,, \\
   {\cal C}_4  &= \frac{(1+z)^3}{{\cal E}^2} {\cal B}_2 \,.     \label{eq:lastObs}
\end{align}
In deriving the above we have defined a quantity that is dependent on the normalization of density fluctuations and, thus, generally not an observable,
\begin{align}
   G &= B_{00}^{\rm sq} - {\cal C}_4^{-2} {\cal B}_1^{-1} \Omega_{\rm m}^2 \Sigma^2 B_0^{\rm llg,sq} \nonumber \\
  &=  ( b_{2,13}^{\rm sq} + b_{2,23}^{\rm sq}) b_1 b_{1,\Delta k} P P_{\Delta k} \,. 
\end{align}

%%%%%%%%%%%%%%%%%%%%%%%%%%%%%%
\section{Evolution equations in squeezed limit}\label{app:squezzed}

In Sec.\,\ref{sec:consistency}, we have derived a relation from the fluid equations by applying the equilateral limit to the wave dependence of the kernels. Here we repeat the analysis for the squeezed case, also to prove that \mbox{${\mathfrak{F}}_{2,12}^{\rm sq} = {\mathfrak{F}}_2(\fett{k},-\fett{k})=0$.}

Using Eqs.\,(\ref{contiF2})--(\ref{eulerG2}) as a starting point and taking the squeezed limit, we obtain, respectively,
\begin{align}
  \left\{ {{\mathfrak{F}}_{2,12}^{\rm sq}}' + 2 f_1 {\mathfrak{F}}_{2,12}^{\rm sq} \right\} \delta_{\rm m1}^2 &=  - {\mathfrak{G}}_{2,12}^{\rm sq}  \delta_{\rm m1}^2 \,, \\
  \left\{ {{\mathfrak{G}}_{2,12}^{\rm sq}}' + 2f_1 {\mathfrak{G}}_{2,12}^{\rm sq} \right\} \delta_{\rm m1}^2 &= \nonumber  \\
 \Bigg\{ - \left( 2+ \frac{H'}{H}\right) {\mathfrak{G}}_{2,12}^{\rm sq}&  - \frac 3 2 \Omega_{\rm m} Y {\mathfrak{F}}_{2,12}^{\rm sq} \Bigg\} \delta_{\rm m1}^2 \,. 
\end{align}
Defining $\delta_2^{\rm sq} = {\mathfrak{F}}_{2,12}^{\rm sq} \delta_{\rm m1}^2$ and 
$\theta_2^{\rm sq} = {\mathfrak{G}}_{2,12}^{\rm sq} \delta_{\rm m1}^2$, we can combine these equations into the following PDE,
\begin{equation}
 {\delta_2^{\rm sq}}'' +   \left( 2+ \frac{H'}{H}\right)  {\delta_2^{\rm sq}}' 
   - \frac 3 2 \Omega_{\rm m} Y \delta_2^{\rm sq} = 0 \,.
\end{equation}
This PDE coincides exactly with the one obtained for the linear matter density, Eq.\,(\ref{evoLinearDelta}), thus its solution will grow with the same linear amplitude $D$. But since $\delta_2$ is of second order with the fastest growing mode potentially of the order of $D^2$, 
we conclude that the above PDE for $\delta_2^{\rm sq}$ excites nothing more but 
decaying modes, and thus we can set $\delta_2^{\rm sq} =0$, 
from which follows that ${\mathfrak{F}}_{2,12}^{\rm sq}=0$.

%%%%%%%%%%%%%%%%%%%%%%%%%%%%%%
\section{Evolution equations for the matter bispectrum}\label{app:nointegral}

In Sec.\,\ref{sec:nonlinearconstraint} we have determined evolution equations that 
lead subsequently to the constraint equation~(\ref{eq:consistency}). For its 
derivation we have argued that we can drop two loop integrals and a Dirac delta. Here we provide
a more rigorous derivation that obviously leads to the identical final result~(\ref{eq:consistency}). 

Actually, to understand our methodology, it is sufficient to focus on the 
lhs term of Eq.\,(\ref{contiF2}) for which we again restore the double integrals and the Dirac delta. In that term we interchange some dependences 
according to $\fett{k} \to \fett{k}_3$, $\fett{k}_1 \to \fett{k}_4$, $\fett{k}_2 \to \fett{k}_5$,
and write equivalently
\begin{widetext}
\begin{align}
 \int  \frac{\dd^3 \fett{k}_{45}}{(2\pi)^3} \delta_{\rm D}^{(3)}(\fett{k}_3 - \fett{k}_{45}  ) \left\{ {\mathfrak{F}}_2'(\fett{k}_4,\fett{k}_5) + {\mathfrak{F}}_2(\fett{k}_4,\fett{k}_5) \left[ f_1(k_4) + f_1(k_5) \right] \right\} \delta_{\rm m1}(k_4) \,\delta_{\rm m1}(k_5) \,.
\end{align}
Multiplying this by $\delta_{\rm m1}(k_1) \,\delta_{\rm m1}(k_2)$ and taking the correlator of the resulting expression, we have
\begin{align}
 \int  \frac{\dd^3 \fett{k}_{45}}{(2\pi)^3} \delta_{\rm D}^{(3)}(\fett{k}_3 - \fett{k}_{45}  ) \left\{ {\mathfrak{F}}_2'(\fett{k}_4,\fett{k}_5) + {\mathfrak{F}}_2(\fett{k}_4,\fett{k}_5) \left[ f_1(k_4) + f_1(k_5) \right] \right\} \big\langle \delta_{\rm m1}(k_1) \,\delta_{\rm m1}(k_2)\, \delta_{\rm m1}(k_4)\, \delta_{\rm m1}(k_5) \big\rangle_{\rm c} \nonumber \\
  = 2 (2\pi)^3 \delta_{\rm D}^{(3)}(\fett{k}_{123} ) \left\{ {\mathfrak{F}}_2'(\fett{k}_1,\fett{k}_2) + {\mathfrak{F}}_2(\fett{k}_1,\fett{k}_2) \left[ f_1(k_1) + f_1(k_2) \right] \right\} P_{\rm m1}(k_1) \,P_{\rm m1}(k_2) \,, \label{closure}
\end{align}
\end{widetext}
where we have used Wick's theorem \cite{Bernardeau:2001qr} and discarded a zero-mode term $\propto \delta_{\rm D}^{(3)}(\fett{k}_3)$. The rhs term in Eq.\,(\ref{closure}) is only nonzero if the closure condition, dictated by $\delta_{\rm D}^{(3)}(\fett{k}_{123} )$, is satisfied. This
is indeed the case for the bispectrum where the three wave vectors form a closed triangle in Fourier space.
By contrast, the omitted zero-mode term that is proportional to $\delta_{\rm D}^{(3)}(\fett{k}_3)$ dictates $\fett{k}_3 =0$ and no triangle closure condition.

The same technique applies to the rhs of Eq.\,(\ref{contiF2}) [and, of course, to the whole Eq.\,(\ref{eulerG2}) as well]; dropping the Dirac delta, some constant factors and the two power spectra, we then obtain for the equilateral triangle configuration Eq.\,(\ref{contiF2eq}) [and Eq.\,(\ref{eulerG2eq}), respectively],
which concludes the proof. 

Finally, we note that the above technique delivers evolution equations not for the fluid variables but for the bispectrum. In general, this technique of course applies not only to the bispectrum but to any polyspectrum.

\

%%%%%%%%%%%%%%%%%%%%%%%%%%%%%%%%%%%%%%%%%%%%%%%%%%%%%%%%%%%%%%%%%%%%%%%%%%%%%%%%%%%%
\bibliography{observables}

\begin{thebibliography}{53}%
\makeatletter
\providecommand \@ifxundefined [1]{%
 \@ifx{#1\undefined}
}%
\providecommand \@ifnum [1]{%
 \ifnum #1\expandafter \@firstoftwo
 \else \expandafter \@secondoftwo
 \fi
}%
\providecommand \@ifx [1]{%
 \ifx #1\expandafter \@firstoftwo
 \else \expandafter \@secondoftwo
 \fi
}%
\providecommand \natexlab [1]{#1}%
\providecommand \enquote  [1]{``#1''}%
\providecommand \bibnamefont  [1]{#1}%
\providecommand \bibfnamefont [1]{#1}%
\providecommand \citenamefont [1]{#1}%
\providecommand \href@noop [0]{\@secondoftwo}%
\providecommand \href [0]{\begingroup \@sanitize@url \@href}%
\providecommand \@href[1]{\@@startlink{#1}\@@href}%
\providecommand \@@href[1]{\endgroup#1\@@endlink}%
\providecommand \@sanitize@url [0]{\catcode `\\12\catcode `\$12\catcode
  `\&12\catcode `\#12\catcode `\^12\catcode `\_12\catcode `\%12\relax}%
\providecommand \@@startlink[1]{}%
\providecommand \@@endlink[0]{}%
\providecommand \url  [0]{\begingroup\@sanitize@url \@url }%
\providecommand \@url [1]{\endgroup\@href {#1}{\urlprefix }}%
\providecommand \urlprefix  [0]{URL }%
\providecommand \Eprint [0]{\href }%
\providecommand \doibase [0]{http://dx.doi.org/}%
\providecommand \selectlanguage [0]{\@gobble}%
\providecommand \bibinfo  [0]{\@secondoftwo}%
\providecommand \bibfield  [0]{\@secondoftwo}%
\providecommand \translation [1]{[#1]}%
\providecommand \BibitemOpen [0]{}%
\providecommand \bibitemStop [0]{}%
\providecommand \bibitemNoStop [0]{.\EOS\space}%
\providecommand \EOS [0]{\spacefactor3000\relax}%
\providecommand \BibitemShut  [1]{\csname bibitem#1\endcsname}%
\let\auto@bib@innerbib\@empty
%</preamble>
\bibitem [{\citenamefont {Desjacques}\ \emph {et~al.}(2016)\citenamefont
  {Desjacques}, \citenamefont {Jeong},\ and\ \citenamefont
  {Schmidt}}]{Desjacques:2016bnm}%
  \BibitemOpen
  \bibfield  {author} {\bibinfo {author} {\bibfnamefont {V.}~\bibnamefont
  {Desjacques}}, \bibinfo {author} {\bibfnamefont {D.}~\bibnamefont {Jeong}}, \
  and\ \bibinfo {author} {\bibfnamefont {F.}~\bibnamefont {Schmidt}},\
  }\href@noop {} {\  (\bibinfo {year} {2016})},\ \Eprint
  {http://arxiv.org/abs/1611.09787} {arXiv:1611.09787 [astro-ph.CO]}
  \BibitemShut {NoStop}%
%%CITATION = ARXIV:1611.09787;%%
\bibitem [{\citenamefont {Amendola}\ and\ \citenamefont
  {Tsujikawa}(2010)}]{Amendola}%
  \BibitemOpen
  \bibfield  {author} {\bibinfo {author} {\bibfnamefont {L.}~\bibnamefont
  {Amendola}}\ and\ \bibinfo {author} {\bibfnamefont {S.}~\bibnamefont
  {Tsujikawa}},\ }\href {\doibase 10.1017/CBO9780511750823} {\emph {\bibinfo
  {title} {{Dark Energy: Theory and observations}}}}\ (\bibinfo  {publisher}
  {Cambridge University Press},\ \bibinfo {address} {Cambridge},\ \bibinfo
  {year} {2010})\BibitemShut {NoStop}%
\bibitem [{\citenamefont {Clifton}\ \emph {et~al.}(2012)\citenamefont
  {Clifton}, \citenamefont {Ferreira}, \citenamefont {Padilla},\ and\
  \citenamefont {Skordis}}]{Clifton:2011jh}%
  \BibitemOpen
  \bibfield  {author} {\bibinfo {author} {\bibfnamefont {T.}~\bibnamefont
  {Clifton}}, \bibinfo {author} {\bibfnamefont {P.~G.}\ \bibnamefont
  {Ferreira}}, \bibinfo {author} {\bibfnamefont {A.}~\bibnamefont {Padilla}}, \
  and\ \bibinfo {author} {\bibfnamefont {C.}~\bibnamefont {Skordis}},\ }\href
  {\doibase 10.1016/j.physrep.2012.01.001} {\bibfield  {journal} {\bibinfo
  {journal} {Phys. Rept.}\ }\textbf {\bibinfo {volume} {513}},\ \bibinfo
  {pages} {1} (\bibinfo {year} {2012})},\ \Eprint
  {http://arxiv.org/abs/1106.2476} {arXiv:1106.2476 [astro-ph.CO]} \BibitemShut
  {NoStop}%
%%CITATION = ARXIV:1106.2476;%%
\bibitem [{\citenamefont {Kunz}(2012)}]{Kunz:2012aw}%
  \BibitemOpen
  \bibfield  {author} {\bibinfo {author} {\bibfnamefont {M.}~\bibnamefont
  {Kunz}},\ }\href {\doibase 10.1016/j.crhy.2012.04.007} {\bibfield  {journal}
  {\bibinfo  {journal} {Comptes Rendus Physique}\ }\textbf {\bibinfo {volume}
  {13}},\ \bibinfo {pages} {539} (\bibinfo {year} {2012})},\ \Eprint
  {http://arxiv.org/abs/1204.5482} {arXiv:1204.5482 [astro-ph.CO]} \BibitemShut
  {NoStop}%
%%CITATION = ARXIV:1204.5482;%%
\bibitem [{\citenamefont {Li}\ and\ \citenamefont
  {Efstathiou}(2012)}]{Li:2011qda}%
  \BibitemOpen
  \bibfield  {author} {\bibinfo {author} {\bibfnamefont {B.}~\bibnamefont
  {Li}}\ and\ \bibinfo {author} {\bibfnamefont {G.}~\bibnamefont
  {Efstathiou}},\ }\href {\doibase 10.1111/j.1365-2966.2011.20404.x} {\bibfield
   {journal} {\bibinfo  {journal} {Mon. Not. Roy. Astron. Soc.}\ }\textbf
  {\bibinfo {volume} {421}},\ \bibinfo {pages} {1431} (\bibinfo {year}
  {2012})},\ \Eprint {http://arxiv.org/abs/1110.6440} {arXiv:1110.6440
  [astro-ph.CO]} \BibitemShut {NoStop}%
%%CITATION = ARXIV:1110.6440;%%
\bibitem [{\citenamefont {Brax}\ \emph {et~al.}(2012)\citenamefont {Brax},
  \citenamefont {Davis},\ and\ \citenamefont {Li}}]{Brax:2011aw}%
  \BibitemOpen
  \bibfield  {author} {\bibinfo {author} {\bibfnamefont {P.}~\bibnamefont
  {Brax}}, \bibinfo {author} {\bibfnamefont {A.-C.}\ \bibnamefont {Davis}}, \
  and\ \bibinfo {author} {\bibfnamefont {B.}~\bibnamefont {Li}},\ }\href
  {\doibase 10.1016/j.physletb.2012.08.002} {\bibfield  {journal} {\bibinfo
  {journal} {Phys. Lett.}\ }\textbf {\bibinfo {volume} {B715}},\ \bibinfo
  {pages} {38} (\bibinfo {year} {2012})},\ \Eprint
  {http://arxiv.org/abs/1111.6613} {arXiv:1111.6613 [astro-ph.CO]} \BibitemShut
  {NoStop}%
%%CITATION = ARXIV:1111.6613;%%
\bibitem [{\citenamefont {Barreira}\ \emph {et~al.}(2012)\citenamefont
  {Barreira}, \citenamefont {Li}, \citenamefont {Baugh},\ and\ \citenamefont
  {Pascoli}}]{Barreira:2012kk}%
  \BibitemOpen
  \bibfield  {author} {\bibinfo {author} {\bibfnamefont {A.}~\bibnamefont
  {Barreira}}, \bibinfo {author} {\bibfnamefont {B.}~\bibnamefont {Li}},
  \bibinfo {author} {\bibfnamefont {C.~M.}\ \bibnamefont {Baugh}}, \ and\
  \bibinfo {author} {\bibfnamefont {S.}~\bibnamefont {Pascoli}},\ }\href
  {\doibase 10.1103/PhysRevD.86.124016} {\bibfield  {journal} {\bibinfo
  {journal} {Phys. Rev.}\ }\textbf {\bibinfo {volume} {D86}},\ \bibinfo {pages}
  {124016} (\bibinfo {year} {2012})},\ \Eprint {http://arxiv.org/abs/1208.0600}
  {arXiv:1208.0600 [astro-ph.CO]} \BibitemShut {NoStop}%
%%CITATION = ARXIV:1208.0600;%%
\bibitem [{\citenamefont {Li}\ \emph {et~al.}(2013)\citenamefont {Li},
  \citenamefont {Hellwing}, \citenamefont {Koyama}, \citenamefont {Zhao},
  \citenamefont {Jennings},\ and\ \citenamefont {Baugh}}]{Li:2012by}%
  \BibitemOpen
  \bibfield  {author} {\bibinfo {author} {\bibfnamefont {B.}~\bibnamefont
  {Li}}, \bibinfo {author} {\bibfnamefont {W.~A.}\ \bibnamefont {Hellwing}},
  \bibinfo {author} {\bibfnamefont {K.}~\bibnamefont {Koyama}}, \bibinfo
  {author} {\bibfnamefont {G.-B.}\ \bibnamefont {Zhao}}, \bibinfo {author}
  {\bibfnamefont {E.}~\bibnamefont {Jennings}}, \ and\ \bibinfo {author}
  {\bibfnamefont {C.~M.}\ \bibnamefont {Baugh}},\ }\href {\doibase
  10.1093/mnras/sts072} {\bibfield  {journal} {\bibinfo  {journal} {Mon. Not.
  Roy. Astron. Soc.}\ }\textbf {\bibinfo {volume} {428}},\ \bibinfo {pages}
  {743} (\bibinfo {year} {2013})},\ \Eprint {http://arxiv.org/abs/1206.4317}
  {arXiv:1206.4317 [astro-ph.CO]} \BibitemShut {NoStop}%
%%CITATION = ARXIV:1206.4317;%%
\bibitem [{\citenamefont {Jennings}\ \emph {et~al.}(2012)\citenamefont
  {Jennings}, \citenamefont {Baugh}, \citenamefont {Li}, \citenamefont {Zhao},\
  and\ \citenamefont {Koyama}}]{Jennings:2012pt}%
  \BibitemOpen
  \bibfield  {author} {\bibinfo {author} {\bibfnamefont {E.}~\bibnamefont
  {Jennings}}, \bibinfo {author} {\bibfnamefont {C.~M.}\ \bibnamefont {Baugh}},
  \bibinfo {author} {\bibfnamefont {B.}~\bibnamefont {Li}}, \bibinfo {author}
  {\bibfnamefont {G.-B.}\ \bibnamefont {Zhao}}, \ and\ \bibinfo {author}
  {\bibfnamefont {K.}~\bibnamefont {Koyama}},\ }\href {\doibase
  10.1111/j.1365-2966.2012.21567.x} {\bibfield  {journal} {\bibinfo  {journal}
  {Mon. Not. Roy. Astron. Soc.}\ }\textbf {\bibinfo {volume} {425}},\ \bibinfo
  {pages} {2128} (\bibinfo {year} {2012})},\ \Eprint
  {http://arxiv.org/abs/1205.2698} {arXiv:1205.2698 [astro-ph.CO]} \BibitemShut
  {NoStop}%
%%CITATION = ARXIV:1205.2698;%%
\bibitem [{\citenamefont {Joudaki}(2012)}]{Joudaki:2012bcf}%
  \BibitemOpen
  \bibfield  {author} {\bibinfo {author} {\bibfnamefont {S.}~\bibnamefont
  {Joudaki}},\ }\emph {\bibinfo {title} {{Beyond the Standard Model of
  Cosmology: Dark Energy, Massive Neutrinos, and Primordial
  Non-Gaussianity}}},\ \href {http://search.proquest.com/docview/1027427067}
  {Ph.D. thesis},\ \bibinfo  {school} {UC, Irvine} (\bibinfo {year}
  {2012})\BibitemShut {NoStop}%
%%CITATION = INSPIRE-1474640;%%
\bibitem [{\citenamefont {Barreira}\ \emph {et~al.}(2013)\citenamefont
  {Barreira}, \citenamefont {Li}, \citenamefont {Sanchez}, \citenamefont
  {Baugh},\ and\ \citenamefont {Pascoli}}]{Barreira:2013jma}%
  \BibitemOpen
  \bibfield  {author} {\bibinfo {author} {\bibfnamefont {A.}~\bibnamefont
  {Barreira}}, \bibinfo {author} {\bibfnamefont {B.}~\bibnamefont {Li}},
  \bibinfo {author} {\bibfnamefont {A.}~\bibnamefont {Sanchez}}, \bibinfo
  {author} {\bibfnamefont {C.~M.}\ \bibnamefont {Baugh}}, \ and\ \bibinfo
  {author} {\bibfnamefont {S.}~\bibnamefont {Pascoli}},\ }\href {\doibase
  10.1103/PhysRevD.87.103511} {\bibfield  {journal} {\bibinfo  {journal} {Phys.
  Rev.}\ }\textbf {\bibinfo {volume} {D87}},\ \bibinfo {pages} {103511}
  (\bibinfo {year} {2013})},\ \Eprint {http://arxiv.org/abs/1302.6241}
  {arXiv:1302.6241 [astro-ph.CO]} \BibitemShut {NoStop}%
%%CITATION = ARXIV:1302.6241;%%
\bibitem [{\citenamefont {Bellini}\ \emph {et~al.}(2015)\citenamefont
  {Bellini}, \citenamefont {Jimenez},\ and\ \citenamefont
  {Verde}}]{Bellini:2015wfa}%
  \BibitemOpen
  \bibfield  {author} {\bibinfo {author} {\bibfnamefont {E.}~\bibnamefont
  {Bellini}}, \bibinfo {author} {\bibfnamefont {R.}~\bibnamefont {Jimenez}}, \
  and\ \bibinfo {author} {\bibfnamefont {L.}~\bibnamefont {Verde}},\ }\href
  {\doibase 10.1088/1475-7516/2015/05/057} {\bibfield  {journal} {\bibinfo
  {journal} {JCAP}\ }\textbf {\bibinfo {volume} {1505}},\ \bibinfo {pages}
  {057} (\bibinfo {year} {2015})},\ \Eprint {http://arxiv.org/abs/1504.04341}
  {arXiv:1504.04341 [astro-ph.CO]} \BibitemShut {NoStop}%
%%CITATION = ARXIV:1504.04341;%%
\bibitem [{\citenamefont {Huterer}\ and\ \citenamefont
  {Starkman}(2003)}]{Huterer:2002hy}%
  \BibitemOpen
  \bibfield  {author} {\bibinfo {author} {\bibfnamefont {D.}~\bibnamefont
  {Huterer}}\ and\ \bibinfo {author} {\bibfnamefont {G.}~\bibnamefont
  {Starkman}},\ }\href {\doibase 10.1103/PhysRevLett.90.031301} {\bibfield
  {journal} {\bibinfo  {journal} {Phys. Rev. Lett.}\ }\textbf {\bibinfo
  {volume} {90}},\ \bibinfo {pages} {031301} (\bibinfo {year} {2003})},\
  \Eprint {http://arxiv.org/abs/astro-ph/0207517} {arXiv:astro-ph/0207517
  [astro-ph]} \BibitemShut {NoStop}%
%%CITATION = ASTRO-PH/0207517;%%
\bibitem [{\citenamefont {Zhang}\ \emph {et~al.}(2007)\citenamefont {Zhang},
  \citenamefont {Liguori}, \citenamefont {Bean},\ and\ \citenamefont
  {Dodelson}}]{Zhang:2007nk}%
  \BibitemOpen
  \bibfield  {author} {\bibinfo {author} {\bibfnamefont {P.}~\bibnamefont
  {Zhang}}, \bibinfo {author} {\bibfnamefont {M.}~\bibnamefont {Liguori}},
  \bibinfo {author} {\bibfnamefont {R.}~\bibnamefont {Bean}}, \ and\ \bibinfo
  {author} {\bibfnamefont {S.}~\bibnamefont {Dodelson}},\ }\href {\doibase
  10.1103/PhysRevLett.99.141302} {\bibfield  {journal} {\bibinfo  {journal}
  {Phys. Rev. Lett.}\ }\textbf {\bibinfo {volume} {99}},\ \bibinfo {pages}
  {141302} (\bibinfo {year} {2007})},\ \Eprint {http://arxiv.org/abs/0704.1932}
  {arXiv:0704.1932 [astro-ph]} \BibitemShut {NoStop}%
%%CITATION = ARXIV:0704.1932;%%
\bibitem [{\citenamefont {Shafieloo}\ and\ \citenamefont
  {Clarkson}(2010)}]{Shafieloo:2009hi}%
  \BibitemOpen
  \bibfield  {author} {\bibinfo {author} {\bibfnamefont {A.}~\bibnamefont
  {Shafieloo}}\ and\ \bibinfo {author} {\bibfnamefont {C.}~\bibnamefont
  {Clarkson}},\ }\href {\doibase 10.1103/PhysRevD.81.083537} {\bibfield
  {journal} {\bibinfo  {journal} {Phys. Rev.}\ }\textbf {\bibinfo {volume}
  {D81}},\ \bibinfo {pages} {083537} (\bibinfo {year} {2010})},\ \Eprint
  {http://arxiv.org/abs/0911.4858} {arXiv:0911.4858 [astro-ph.CO]} \BibitemShut
  {NoStop}%
%%CITATION = ARXIV:0911.4858;%%
\bibitem [{\citenamefont {Zhao}\ \emph {et~al.}(2009)\citenamefont {Zhao},
  \citenamefont {Pogosian}, \citenamefont {Silvestri},\ and\ \citenamefont
  {Zylberberg}}]{Zhao:2009fn}%
  \BibitemOpen
  \bibfield  {author} {\bibinfo {author} {\bibfnamefont {G.-B.}\ \bibnamefont
  {Zhao}}, \bibinfo {author} {\bibfnamefont {L.}~\bibnamefont {Pogosian}},
  \bibinfo {author} {\bibfnamefont {A.}~\bibnamefont {Silvestri}}, \ and\
  \bibinfo {author} {\bibfnamefont {J.}~\bibnamefont {Zylberberg}},\ }\href
  {\doibase 10.1103/PhysRevLett.103.241301} {\bibfield  {journal} {\bibinfo
  {journal} {Phys. Rev. Lett.}\ }\textbf {\bibinfo {volume} {103}},\ \bibinfo
  {pages} {241301} (\bibinfo {year} {2009})},\ \Eprint
  {http://arxiv.org/abs/0905.1326} {arXiv:0905.1326 [astro-ph.CO]} \BibitemShut
  {NoStop}%
%%CITATION = ARXIV:0905.1326;%%
\bibitem [{\citenamefont {Zhao}\ \emph {et~al.}(2010)\citenamefont {Zhao},
  \citenamefont {Giannantonio}, \citenamefont {Pogosian}, \citenamefont
  {Silvestri}, \citenamefont {Bacon}, \citenamefont {Koyama}, \citenamefont
  {Nichol},\ and\ \citenamefont {Song}}]{Zhao:2010dz}%
  \BibitemOpen
  \bibfield  {author} {\bibinfo {author} {\bibfnamefont {G.-B.}\ \bibnamefont
  {Zhao}}, \bibinfo {author} {\bibfnamefont {T.}~\bibnamefont {Giannantonio}},
  \bibinfo {author} {\bibfnamefont {L.}~\bibnamefont {Pogosian}}, \bibinfo
  {author} {\bibfnamefont {A.}~\bibnamefont {Silvestri}}, \bibinfo {author}
  {\bibfnamefont {D.~J.}\ \bibnamefont {Bacon}}, \bibinfo {author}
  {\bibfnamefont {K.}~\bibnamefont {Koyama}}, \bibinfo {author} {\bibfnamefont
  {R.~C.}\ \bibnamefont {Nichol}}, \ and\ \bibinfo {author} {\bibfnamefont
  {Y.-S.}\ \bibnamefont {Song}},\ }\href {\doibase 10.1103/PhysRevD.81.103510}
  {\bibfield  {journal} {\bibinfo  {journal} {Phys. Rev.}\ }\textbf {\bibinfo
  {volume} {D81}},\ \bibinfo {pages} {103510} (\bibinfo {year} {2010})},\
  \Eprint {http://arxiv.org/abs/1003.0001} {arXiv:1003.0001 [astro-ph.CO]}
  \BibitemShut {NoStop}%
%%CITATION = ARXIV:1003.0001;%%
\bibitem [{\citenamefont {Shafieloo}\ and\ \citenamefont
  {Linder}(2011)}]{Shafieloo:2011zv}%
  \BibitemOpen
  \bibfield  {author} {\bibinfo {author} {\bibfnamefont {A.}~\bibnamefont
  {Shafieloo}}\ and\ \bibinfo {author} {\bibfnamefont {E.~V.}\ \bibnamefont
  {Linder}},\ }\href {\doibase 10.1103/PhysRevD.84.063519} {\bibfield
  {journal} {\bibinfo  {journal} {Phys. Rev.}\ }\textbf {\bibinfo {volume}
  {D84}},\ \bibinfo {pages} {063519} (\bibinfo {year} {2011})},\ \Eprint
  {http://arxiv.org/abs/1107.1033} {arXiv:1107.1033 [astro-ph.CO]} \BibitemShut
  {NoStop}%
%%CITATION = ARXIV:1107.1033;%%
\bibitem [{\citenamefont {Stebbins}(2012)}]{Stebbins:2012vw}%
  \BibitemOpen
  \bibfield  {author} {\bibinfo {author} {\bibfnamefont {A.}~\bibnamefont
  {Stebbins}},\ }\href {\doibase 10.1142/S0218271812420175} {\bibfield
  {journal} {\bibinfo  {journal} {Int. J. Mod. Phys.}\ }\textbf {\bibinfo
  {volume} {D21}},\ \bibinfo {pages} {1242017} (\bibinfo {year} {2012})},\
  \Eprint {http://arxiv.org/abs/1205.4201} {arXiv:1205.4201 [gr-qc]}
  \BibitemShut {NoStop}%
%%CITATION = ARXIV:1205.4201;%%
\bibitem [{\citenamefont {Valkenburg}\ \emph {et~al.}(2014)\citenamefont
  {Valkenburg}, \citenamefont {Marra},\ and\ \citenamefont
  {Clarkson}}]{Valkenburg:2012td}%
  \BibitemOpen
  \bibfield  {author} {\bibinfo {author} {\bibfnamefont {W.}~\bibnamefont
  {Valkenburg}}, \bibinfo {author} {\bibfnamefont {V.}~\bibnamefont {Marra}}, \
  and\ \bibinfo {author} {\bibfnamefont {C.}~\bibnamefont {Clarkson}},\ }\href
  {\doibase 10.1093/mnrasl/slt140} {\bibfield  {journal} {\bibinfo  {journal}
  {Mon. Not. Roy. Astron. Soc.}\ }\textbf {\bibinfo {volume} {438}},\ \bibinfo
  {pages} {L6} (\bibinfo {year} {2014})},\ \Eprint
  {http://arxiv.org/abs/1209.4078} {arXiv:1209.4078 [astro-ph.CO]} \BibitemShut
  {NoStop}%
%%CITATION = ARXIV:1209.4078;%%
\bibitem [{\citenamefont {Amendola}\ \emph {et~al.}(2013)\citenamefont
  {Amendola}, \citenamefont {Kunz}, \citenamefont {Motta}, \citenamefont
  {Saltas},\ and\ \citenamefont {Sawicki}}]{Amendola:2012ky}%
  \BibitemOpen
  \bibfield  {author} {\bibinfo {author} {\bibfnamefont {L.}~\bibnamefont
  {Amendola}}, \bibinfo {author} {\bibfnamefont {M.}~\bibnamefont {Kunz}},
  \bibinfo {author} {\bibfnamefont {M.}~\bibnamefont {Motta}}, \bibinfo
  {author} {\bibfnamefont {I.~D.}\ \bibnamefont {Saltas}}, \ and\ \bibinfo
  {author} {\bibfnamefont {I.}~\bibnamefont {Sawicki}},\ }\href {\doibase
  10.1103/PhysRevD.87.023501} {\bibfield  {journal} {\bibinfo  {journal} {Phys.
  Rev.}\ }\textbf {\bibinfo {volume} {D87}},\ \bibinfo {pages} {023501}
  (\bibinfo {year} {2013})},\ \Eprint {http://arxiv.org/abs/1210.0439}
  {arXiv:1210.0439 [astro-ph.CO]} \BibitemShut {NoStop}%
%%CITATION = ARXIV:1210.0439;%%
\bibitem [{\citenamefont {Baker}\ \emph {et~al.}(2014)\citenamefont {Baker},
  \citenamefont {Ferreira},\ and\ \citenamefont {Skordis}}]{Baker:2013hia}%
  \BibitemOpen
  \bibfield  {author} {\bibinfo {author} {\bibfnamefont {T.}~\bibnamefont
  {Baker}}, \bibinfo {author} {\bibfnamefont {P.~G.}\ \bibnamefont {Ferreira}},
  \ and\ \bibinfo {author} {\bibfnamefont {C.}~\bibnamefont {Skordis}},\ }\href
  {\doibase 10.1103/PhysRevD.89.024026} {\bibfield  {journal} {\bibinfo
  {journal} {Phys. Rev.}\ }\textbf {\bibinfo {volume} {D89}},\ \bibinfo {pages}
  {024026} (\bibinfo {year} {2014})},\ \Eprint {http://arxiv.org/abs/1310.1086}
  {arXiv:1310.1086 [astro-ph.CO]} \BibitemShut {NoStop}%
%%CITATION = ARXIV:1310.1086;%%
\bibitem [{\citenamefont {Alam}\ \emph {et~al.}(2016)\citenamefont {Alam},
  \citenamefont {Ho},\ and\ \citenamefont {Silvestri}}]{Alam:2015rsa}%
  \BibitemOpen
  \bibfield  {author} {\bibinfo {author} {\bibfnamefont {S.}~\bibnamefont
  {Alam}}, \bibinfo {author} {\bibfnamefont {S.}~\bibnamefont {Ho}}, \ and\
  \bibinfo {author} {\bibfnamefont {A.}~\bibnamefont {Silvestri}},\ }\href
  {\doibase 10.1093/mnras/stv2935} {\bibfield  {journal} {\bibinfo  {journal}
  {Mon. Not. Roy. Astron. Soc.}\ }\textbf {\bibinfo {volume} {456}},\ \bibinfo
  {pages} {3743} (\bibinfo {year} {2016})},\ \Eprint
  {http://arxiv.org/abs/1509.05034} {arXiv:1509.05034 [astro-ph.CO]}
  \BibitemShut {NoStop}%
%%CITATION = ARXIV:1509.05034;%%
\bibitem [{\citenamefont {Taddei}\ \emph {et~al.}(2016)\citenamefont {Taddei},
  \citenamefont {Martinelli},\ and\ \citenamefont {Amendola}}]{Taddei:2016iku}%
  \BibitemOpen
  \bibfield  {author} {\bibinfo {author} {\bibfnamefont {L.}~\bibnamefont
  {Taddei}}, \bibinfo {author} {\bibfnamefont {M.}~\bibnamefont {Martinelli}},
  \ and\ \bibinfo {author} {\bibfnamefont {L.}~\bibnamefont {Amendola}},\
  }\href {\doibase 10.1088/1475-7516/2016/12/032} {\bibfield  {journal}
  {\bibinfo  {journal} {JCAP}\ }\textbf {\bibinfo {volume} {1612}},\ \bibinfo
  {pages} {032} (\bibinfo {year} {2016})},\ \Eprint
  {http://arxiv.org/abs/1604.01059} {arXiv:1604.01059 [astro-ph.CO]}
  \BibitemShut {NoStop}%
%%CITATION = ARXIV:1604.01059;%%
\bibitem [{\citenamefont {Netterfield}\ \emph {et~al.}(2002)\citenamefont
  {Netterfield} \emph {et~al.}}]{Netterfield:2001yq}%
  \BibitemOpen
  \bibfield  {author} {\bibinfo {author} {\bibfnamefont {C.~B.}\ \bibnamefont
  {Netterfield}} \emph {et~al.} (\bibinfo {collaboration} {Boomerang}),\ }\href
  {\doibase 10.1086/340118} {\bibfield  {journal} {\bibinfo  {journal}
  {Astrophys. J.}\ }\textbf {\bibinfo {volume} {571}},\ \bibinfo {pages} {604}
  (\bibinfo {year} {2002})},\ \Eprint {http://arxiv.org/abs/astro-ph/0104460}
  {arXiv:astro-ph/0104460 [astro-ph]} \BibitemShut {NoStop}%
%%CITATION = ASTRO-PH/0104460;%%
\bibitem [{\citenamefont {Kunz}(2009)}]{Kunz:2007rk}%
  \BibitemOpen
  \bibfield  {author} {\bibinfo {author} {\bibfnamefont {M.}~\bibnamefont
  {Kunz}},\ }\href {\doibase 10.1103/PhysRevD.80.123001} {\bibfield  {journal}
  {\bibinfo  {journal} {Phys. Rev.}\ }\textbf {\bibinfo {volume} {D80}},\
  \bibinfo {pages} {123001} (\bibinfo {year} {2009})},\ \Eprint
  {http://arxiv.org/abs/astro-ph/0702615} {arXiv:astro-ph/0702615 [astro-ph]}
  \BibitemShut {NoStop}%
%%CITATION = ASTRO-PH/0702615;%%
\bibitem [{\citenamefont {Hellwing}\ \emph {et~al.}(2016)\citenamefont
  {Hellwing}, \citenamefont {Schaller}, \citenamefont {Frenk}, \citenamefont
  {Theuns}, \citenamefont {Schaye}, \citenamefont {Bower},\ and\ \citenamefont
  {Crain}}]{Hellwing:2016ucy}%
  \BibitemOpen
  \bibfield  {author} {\bibinfo {author} {\bibfnamefont {W.~A.}\ \bibnamefont
  {Hellwing}}, \bibinfo {author} {\bibfnamefont {M.}~\bibnamefont {Schaller}},
  \bibinfo {author} {\bibfnamefont {C.~S.}\ \bibnamefont {Frenk}}, \bibinfo
  {author} {\bibfnamefont {T.}~\bibnamefont {Theuns}}, \bibinfo {author}
  {\bibfnamefont {J.}~\bibnamefont {Schaye}}, \bibinfo {author} {\bibfnamefont
  {R.~G.}\ \bibnamefont {Bower}}, \ and\ \bibinfo {author} {\bibfnamefont
  {R.~A.}\ \bibnamefont {Crain}},\ }\href {\doibase 10.1093/mnrasl/slw081}
  {\bibfield  {journal} {\bibinfo  {journal} {Mon. Not. Roy. Astron. Soc.}\
  }\textbf {\bibinfo {volume} {461}},\ \bibinfo {pages} {L11} (\bibinfo {year}
  {2016})},\ \Eprint {http://arxiv.org/abs/1603.03328} {arXiv:1603.03328
  [astro-ph.CO]} \BibitemShut {NoStop}%
%%CITATION = ARXIV:1603.03328;%%
\bibitem [{\citenamefont {Villa}\ and\ \citenamefont
  {Rampf}(2016)}]{Villa:2015ppa}%
  \BibitemOpen
  \bibfield  {author} {\bibinfo {author} {\bibfnamefont {E.}~\bibnamefont
  {Villa}}\ and\ \bibinfo {author} {\bibfnamefont {C.}~\bibnamefont {Rampf}},\
  }\href {\doibase 10.1088/1475-7516/2016/01/030} {\bibfield  {journal}
  {\bibinfo  {journal} {JCAP}\ }\textbf {\bibinfo {volume} {1601}},\ \bibinfo
  {pages} {030} (\bibinfo {year} {2016})},\ \Eprint
  {http://arxiv.org/abs/1505.04782} {arXiv:1505.04782 [gr-qc]} \BibitemShut
  {NoStop}%
%%CITATION = ARXIV:1505.04782;%%
\bibitem [{\citenamefont {Tram}\ \emph {et~al.}(2016)\citenamefont {Tram},
  \citenamefont {Fidler}, \citenamefont {Crittenden}, \citenamefont {Koyama},
  \citenamefont {Pettinari},\ and\ \citenamefont {Wands}}]{Tram:2016cpy}%
  \BibitemOpen
  \bibfield  {author} {\bibinfo {author} {\bibfnamefont {T.}~\bibnamefont
  {Tram}}, \bibinfo {author} {\bibfnamefont {C.}~\bibnamefont {Fidler}},
  \bibinfo {author} {\bibfnamefont {R.}~\bibnamefont {Crittenden}}, \bibinfo
  {author} {\bibfnamefont {K.}~\bibnamefont {Koyama}}, \bibinfo {author}
  {\bibfnamefont {G.~W.}\ \bibnamefont {Pettinari}}, \ and\ \bibinfo {author}
  {\bibfnamefont {D.}~\bibnamefont {Wands}},\ }\href {\doibase
  10.1088/1475-7516/2016/05/058} {\bibfield  {journal} {\bibinfo  {journal}
  {JCAP}\ }\textbf {\bibinfo {volume} {1605}},\ \bibinfo {pages} {058}
  (\bibinfo {year} {2016})},\ \Eprint {http://arxiv.org/abs/1602.05933}
  {arXiv:1602.05933 [astro-ph.CO]} \BibitemShut {NoStop}%
%%CITATION = ARXIV:1602.05933;%%
\bibitem [{\citenamefont {Di~Dio}\ \emph {et~al.}(2017)\citenamefont {Di~Dio},
  \citenamefont {Perrier}, \citenamefont {Durrer}, \citenamefont {Marozzi},
  \citenamefont {Dizgah}, \citenamefont {Norena},\ and\ \citenamefont
  {Riotto}}]{DiDio:2016gpd}%
  \BibitemOpen
  \bibfield  {author} {\bibinfo {author} {\bibfnamefont {E.}~\bibnamefont
  {Di~Dio}}, \bibinfo {author} {\bibfnamefont {H.}~\bibnamefont {Perrier}},
  \bibinfo {author} {\bibfnamefont {R.}~\bibnamefont {Durrer}}, \bibinfo
  {author} {\bibfnamefont {G.}~\bibnamefont {Marozzi}}, \bibinfo {author}
  {\bibfnamefont {A.~M.}\ \bibnamefont {Dizgah}}, \bibinfo {author}
  {\bibfnamefont {J.}~\bibnamefont {Norena}}, \ and\ \bibinfo {author}
  {\bibfnamefont {A.}~\bibnamefont {Riotto}},\ }\href {\doibase
  10.1088/1475-7516/2017/03/006} {\bibfield  {journal} {\bibinfo  {journal}
  {JCAP}\ }\textbf {\bibinfo {volume} {1703}},\ \bibinfo {pages} {006}
  (\bibinfo {year} {2017})},\ \Eprint {http://arxiv.org/abs/1611.03720}
  {arXiv:1611.03720 [astro-ph.CO]} \BibitemShut {NoStop}%
%%CITATION = ARXIV:1611.03720;%%
\bibitem [{\citenamefont {Bernardeau}\ \emph {et~al.}(2002)\citenamefont
  {Bernardeau}, \citenamefont {Colombi}, \citenamefont {Gaztanaga},\ and\
  \citenamefont {Scoccimarro}}]{Bernardeau:2001qr}%
  \BibitemOpen
  \bibfield  {author} {\bibinfo {author} {\bibfnamefont {F.}~\bibnamefont
  {Bernardeau}}, \bibinfo {author} {\bibfnamefont {S.}~\bibnamefont {Colombi}},
  \bibinfo {author} {\bibfnamefont {E.}~\bibnamefont {Gaztanaga}}, \ and\
  \bibinfo {author} {\bibfnamefont {R.}~\bibnamefont {Scoccimarro}},\ }\href
  {\doibase 10.1016/S0370-1573(02)00135-7} {\bibfield  {journal} {\bibinfo
  {journal} {Phys. Rep.}\ }\textbf {\bibinfo {volume} {367}},\ \bibinfo {pages}
  {1} (\bibinfo {year} {2002})},\ \Eprint
  {http://arxiv.org/abs/astro-ph/0112551} {arXiv:astro-ph/0112551 [astro-ph]}
  \BibitemShut {NoStop}%
%%CITATION = ASTRO-PH/0112551;%%
\bibitem [{\citenamefont {Kaiser}(1987)}]{Kaiser:1987qv}%
  \BibitemOpen
  \bibfield  {author} {\bibinfo {author} {\bibfnamefont {N.}~\bibnamefont
  {Kaiser}},\ }\href@noop {} {\bibfield  {journal} {\bibinfo  {journal} {Mon.
  Not. Roy. Astron. Soc.}\ }\textbf {\bibinfo {volume} {227}},\ \bibinfo
  {pages} {1} (\bibinfo {year} {1987})}\BibitemShut {NoStop}%
%%CITATION = MNRAA,227,1;%%
\bibitem [{\citenamefont {Matsubara}(2011)}]{Matsubara:2011ck}%
  \BibitemOpen
  \bibfield  {author} {\bibinfo {author} {\bibfnamefont {T.}~\bibnamefont
  {Matsubara}},\ }\href {\doibase 10.1103/PhysRevD.83.083518} {\bibfield
  {journal} {\bibinfo  {journal} {Phys. Rev.}\ }\textbf {\bibinfo {volume}
  {D83}},\ \bibinfo {pages} {083518} (\bibinfo {year} {2011})},\ \Eprint
  {http://arxiv.org/abs/1102.4619} {arXiv:1102.4619 [astro-ph.CO]} \BibitemShut
  {NoStop}%
%%CITATION = ARXIV:1102.4619;%%
\bibitem [{\citenamefont {Scoccimarro}\ \emph {et~al.}(1999)\citenamefont
  {Scoccimarro}, \citenamefont {Couchman},\ and\ \citenamefont
  {Frieman}}]{Scoccimarro:1999ed}%
  \BibitemOpen
  \bibfield  {author} {\bibinfo {author} {\bibfnamefont {R.}~\bibnamefont
  {Scoccimarro}}, \bibinfo {author} {\bibfnamefont {H.~M.~P.}\ \bibnamefont
  {Couchman}}, \ and\ \bibinfo {author} {\bibfnamefont {J.~A.}\ \bibnamefont
  {Frieman}},\ }\href {\doibase 10.1086/307220} {\bibfield  {journal} {\bibinfo
   {journal} {Astrophys. J.}\ }\textbf {\bibinfo {volume} {517}},\ \bibinfo
  {pages} {531} (\bibinfo {year} {1999})},\ \Eprint
  {http://arxiv.org/abs/astro-ph/9808305} {arXiv:astro-ph/9808305 [astro-ph]}
  \BibitemShut {NoStop}%
%%CITATION = ASTRO-PH/9808305;%%
\bibitem [{\citenamefont {Rampf}\ and\ \citenamefont
  {Wiegand}(2014)}]{Rampf:2014mga}%
  \BibitemOpen
  \bibfield  {author} {\bibinfo {author} {\bibfnamefont {C.}~\bibnamefont
  {Rampf}}\ and\ \bibinfo {author} {\bibfnamefont {A.}~\bibnamefont
  {Wiegand}},\ }\href {\doibase 10.1103/PhysRevD.90.123503} {\bibfield
  {journal} {\bibinfo  {journal} {Phys. Rev.}\ }\textbf {\bibinfo {volume}
  {D90}},\ \bibinfo {pages} {123503} (\bibinfo {year} {2014})},\ \Eprint
  {http://arxiv.org/abs/1409.2688} {arXiv:1409.2688 [gr-qc]} \BibitemShut
  {NoStop}%
%%CITATION = ARXIV:1409.2688;%%
\bibitem [{\citenamefont {Dekel}\ and\ \citenamefont
  {Lahav}(1999)}]{Dekel:1998eq}%
  \BibitemOpen
  \bibfield  {author} {\bibinfo {author} {\bibfnamefont {A.}~\bibnamefont
  {Dekel}}\ and\ \bibinfo {author} {\bibfnamefont {O.}~\bibnamefont {Lahav}},\
  }\href {\doibase 10.1086/307428} {\bibfield  {journal} {\bibinfo  {journal}
  {Astrophys. J.}\ }\textbf {\bibinfo {volume} {520}},\ \bibinfo {pages} {24}
  (\bibinfo {year} {1999})},\ \Eprint {http://arxiv.org/abs/astro-ph/9806193}
  {arXiv:astro-ph/9806193 [astro-ph]} \BibitemShut {NoStop}%
%%CITATION = ASTRO-PH/9806193;%%
\bibitem [{\citenamefont {Dodelson}(2003)}]{Dodelson:2003ft}%
  \BibitemOpen
  \bibfield  {author} {\bibinfo {author} {\bibfnamefont {S.}~\bibnamefont
  {Dodelson}},\ }\href
  {http://www.slac.stanford.edu/spires/find/books/www?cl=QB981:D62:2003} {\emph
  {\bibinfo {title} {{Modern Cosmology}}}}\ (\bibinfo  {publisher} {Academic
  Press},\ \bibinfo {address} {Amsterdam},\ \bibinfo {year} {2003})\BibitemShut
  {NoStop}%
%%CITATION = INSPIRE-640063;%%
\bibitem [{\citenamefont {Percival}\ and\ \citenamefont
  {White}(2009)}]{Percival:2008sh}%
  \BibitemOpen
  \bibfield  {author} {\bibinfo {author} {\bibfnamefont {W.~J.}\ \bibnamefont
  {Percival}}\ and\ \bibinfo {author} {\bibfnamefont {M.}~\bibnamefont
  {White}},\ }\href {\doibase 10.1111/j.1365-2966.2008.14211.x} {\bibfield
  {journal} {\bibinfo  {journal} {Mon. Not. Roy. Astron. Soc.}\ }\textbf
  {\bibinfo {volume} {393}},\ \bibinfo {pages} {297} (\bibinfo {year}
  {2009})},\ \Eprint {http://arxiv.org/abs/0808.0003} {arXiv:0808.0003
  [astro-ph]} \BibitemShut {NoStop}%
%%CITATION = ARXIV:0808.0003;%%
\bibitem [{\citenamefont {Cole}\ \emph {et~al.}(1994)\citenamefont {Cole},
  \citenamefont {Fisher},\ and\ \citenamefont {Weinberg}}]{Cole:1993kh}%
  \BibitemOpen
  \bibfield  {author} {\bibinfo {author} {\bibfnamefont {S.}~\bibnamefont
  {Cole}}, \bibinfo {author} {\bibfnamefont {K.~B.}\ \bibnamefont {Fisher}}, \
  and\ \bibinfo {author} {\bibfnamefont {D.~H.}\ \bibnamefont {Weinberg}},\
  }\href {\doibase 10.1093/mnras/267.3.785} {\bibfield  {journal} {\bibinfo
  {journal} {Mon. Not. Roy. Astron. Soc.}\ }\textbf {\bibinfo {volume} {267}},\
  \bibinfo {pages} {785} (\bibinfo {year} {1994})},\ \Eprint
  {http://arxiv.org/abs/astro-ph/9308003} {arXiv:astro-ph/9308003 [astro-ph]}
  \BibitemShut {NoStop}%
%%CITATION = ASTRO-PH/9308003;%%
\bibitem [{\citenamefont {Hatton}\ and\ \citenamefont
  {Cole}(1998)}]{Hatton:1997xs}%
  \BibitemOpen
  \bibfield  {author} {\bibinfo {author} {\bibfnamefont {S.~J.}\ \bibnamefont
  {Hatton}}\ and\ \bibinfo {author} {\bibfnamefont {S.}~\bibnamefont {Cole}},\
  }\href {\doibase 10.1046/j.1365-8711.1998.01269.x} {\bibfield  {journal}
  {\bibinfo  {journal} {Mon. Not. Roy. Astron. Soc.}\ }\textbf {\bibinfo
  {volume} {296}},\ \bibinfo {pages} {10} (\bibinfo {year} {1998})},\ \Eprint
  {http://arxiv.org/abs/astro-ph/9707186} {arXiv:astro-ph/9707186 [astro-ph]}
  \BibitemShut {NoStop}%
%%CITATION = ASTRO-PH/9707186;%%
\bibitem [{\citenamefont {Pueblas}\ and\ \citenamefont
  {Scoccimarro}(2009)}]{Pueblas:2008uv}%
  \BibitemOpen
  \bibfield  {author} {\bibinfo {author} {\bibfnamefont {S.}~\bibnamefont
  {Pueblas}}\ and\ \bibinfo {author} {\bibfnamefont {R.}~\bibnamefont
  {Scoccimarro}},\ }\href {\doibase 10.1103/PhysRevD.80.043504} {\bibfield
  {journal} {\bibinfo  {journal} {Phys. Rev.}\ }\textbf {\bibinfo {volume}
  {D80}},\ \bibinfo {pages} {043504} (\bibinfo {year} {2009})},\ \Eprint
  {http://arxiv.org/abs/0809.4606} {arXiv:0809.4606 [astro-ph]} \BibitemShut
  {NoStop}%
%%CITATION = ARXIV:0809.4606;%%
\bibitem [{\citenamefont {Cusin}\ \emph {et~al.}(2017)\citenamefont {Cusin},
  \citenamefont {Tansella},\ and\ \citenamefont {Durrer}}]{Cusin:2016zvu}%
  \BibitemOpen
  \bibfield  {author} {\bibinfo {author} {\bibfnamefont {G.}~\bibnamefont
  {Cusin}}, \bibinfo {author} {\bibfnamefont {V.}~\bibnamefont {Tansella}}, \
  and\ \bibinfo {author} {\bibfnamefont {R.}~\bibnamefont {Durrer}},\ }\href
  {\doibase 10.1103/PhysRevD.95.063527} {\bibfield  {journal} {\bibinfo
  {journal} {Phys. Rev.}\ }\textbf {\bibinfo {volume} {D95}},\ \bibinfo {pages}
  {063527} (\bibinfo {year} {2017})},\ \Eprint
  {http://arxiv.org/abs/1612.00783} {arXiv:1612.00783 [astro-ph.CO]}
  \BibitemShut {NoStop}%
%%CITATION = ARXIV:1612.00783;%%
\bibitem [{\citenamefont {Matsubara}(2008)}]{Matsubara:2007wj}%
  \BibitemOpen
  \bibfield  {author} {\bibinfo {author} {\bibfnamefont {T.}~\bibnamefont
  {Matsubara}},\ }\href {\doibase 10.1103/PhysRevD.77.063530} {\bibfield
  {journal} {\bibinfo  {journal} {Phys. Rev.}\ }\textbf {\bibinfo {volume}
  {D77}},\ \bibinfo {pages} {063530} (\bibinfo {year} {2008})},\ \Eprint
  {http://arxiv.org/abs/0711.2521} {arXiv:0711.2521 [astro-ph]} \BibitemShut
  {NoStop}%
%%CITATION = ARXIV:0711.2521;%%
\bibitem [{\citenamefont {Rampf}\ and\ \citenamefont
  {Wong}(2012)}]{Rampf:2012xb}%
  \BibitemOpen
  \bibfield  {author} {\bibinfo {author} {\bibfnamefont {C.}~\bibnamefont
  {Rampf}}\ and\ \bibinfo {author} {\bibfnamefont {Y.~Y.~Y.}\ \bibnamefont
  {Wong}},\ }\href {\doibase 10.1088/1475-7516/2012/06/018} {\bibfield
  {journal} {\bibinfo  {journal} {JCAP}\ }\textbf {\bibinfo {volume} {1206}},\
  \bibinfo {pages} {018} (\bibinfo {year} {2012})},\ \Eprint
  {http://arxiv.org/abs/1203.4261} {arXiv:1203.4261 [astro-ph.CO]} \BibitemShut
  {NoStop}%
%%CITATION = ARXIV:1203.4261;%%
\bibitem [{\citenamefont {Seljak}\ and\ \citenamefont
  {McDonald}(2011)}]{Seljak:2011tx}%
  \BibitemOpen
  \bibfield  {author} {\bibinfo {author} {\bibfnamefont {U.}~\bibnamefont
  {Seljak}}\ and\ \bibinfo {author} {\bibfnamefont {P.}~\bibnamefont
  {McDonald}},\ }\href {\doibase 10.1088/1475-7516/2011/11/039} {\bibfield
  {journal} {\bibinfo  {journal} {JCAP}\ }\textbf {\bibinfo {volume} {1111}},\
  \bibinfo {pages} {039} (\bibinfo {year} {2011})},\ \Eprint
  {http://arxiv.org/abs/1109.1888} {arXiv:1109.1888 [astro-ph.CO]} \BibitemShut
  {NoStop}%
%%CITATION = ARXIV:1109.1888;%%
\bibitem [{\citenamefont {Vlah}\ \emph {et~al.}(2012)\citenamefont {Vlah},
  \citenamefont {Seljak}, \citenamefont {McDonald}, \citenamefont {Okumura},\
  and\ \citenamefont {Baldauf}}]{Vlah:2012ni}%
  \BibitemOpen
  \bibfield  {author} {\bibinfo {author} {\bibfnamefont {Z.}~\bibnamefont
  {Vlah}}, \bibinfo {author} {\bibfnamefont {U.}~\bibnamefont {Seljak}},
  \bibinfo {author} {\bibfnamefont {P.}~\bibnamefont {McDonald}}, \bibinfo
  {author} {\bibfnamefont {T.}~\bibnamefont {Okumura}}, \ and\ \bibinfo
  {author} {\bibfnamefont {T.}~\bibnamefont {Baldauf}},\ }\href {\doibase
  10.1088/1475-7516/2012/11/009} {\bibfield  {journal} {\bibinfo  {journal}
  {JCAP}\ }\textbf {\bibinfo {volume} {1211}},\ \bibinfo {pages} {009}
  (\bibinfo {year} {2012})},\ \Eprint {http://arxiv.org/abs/1207.0839}
  {arXiv:1207.0839 [astro-ph.CO]} \BibitemShut {NoStop}%
%%CITATION = ARXIV:1207.0839;%%
\bibitem [{\citenamefont {Peacock}\ and\ \citenamefont
  {Dodds}(1994)}]{Peacock:1993xg}%
  \BibitemOpen
  \bibfield  {author} {\bibinfo {author} {\bibfnamefont {J.~A.}\ \bibnamefont
  {Peacock}}\ and\ \bibinfo {author} {\bibfnamefont {S.~J.}\ \bibnamefont
  {Dodds}},\ }\href {\doibase 10.1093/mnras/267.4.1020} {\bibfield  {journal}
  {\bibinfo  {journal} {Mon. Not. Roy. Astron. Soc.}\ }\textbf {\bibinfo
  {volume} {267}},\ \bibinfo {pages} {1020} (\bibinfo {year} {1994})},\ \Eprint
  {http://arxiv.org/abs/astro-ph/9311057} {arXiv:astro-ph/9311057 [astro-ph]}
  \BibitemShut {NoStop}%
%%CITATION = ASTRO-PH/9311057;%%
\bibitem [{\citenamefont {Taruya}\ \emph {et~al.}(2010)\citenamefont {Taruya},
  \citenamefont {Nishimichi},\ and\ \citenamefont {Saito}}]{Taruya:2010mx}%
  \BibitemOpen
  \bibfield  {author} {\bibinfo {author} {\bibfnamefont {A.}~\bibnamefont
  {Taruya}}, \bibinfo {author} {\bibfnamefont {T.}~\bibnamefont {Nishimichi}},
  \ and\ \bibinfo {author} {\bibfnamefont {S.}~\bibnamefont {Saito}},\ }\href
  {\doibase 10.1103/PhysRevD.82.063522} {\bibfield  {journal} {\bibinfo
  {journal} {Phys. Rev.}\ }\textbf {\bibinfo {volume} {D82}},\ \bibinfo {pages}
  {063522} (\bibinfo {year} {2010})},\ \Eprint {http://arxiv.org/abs/1006.0699}
  {arXiv:1006.0699 [astro-ph.CO]} \BibitemShut {NoStop}%
%%CITATION = ARXIV:1006.0699;%%
\bibitem [{\citenamefont {Hashimoto}\ \emph {et~al.}(2017)\citenamefont
  {Hashimoto}, \citenamefont {Rasera},\ and\ \citenamefont
  {Taruya}}]{Hashimoto:2017klo}%
  \BibitemOpen
  \bibfield  {author} {\bibinfo {author} {\bibfnamefont {I.}~\bibnamefont
  {Hashimoto}}, \bibinfo {author} {\bibfnamefont {Y.}~\bibnamefont {Rasera}}, \
  and\ \bibinfo {author} {\bibfnamefont {A.}~\bibnamefont {Taruya}},\
  }\href@noop {} {\  (\bibinfo {year} {2017})},\ \Eprint
  {http://arxiv.org/abs/1705.02574} {arXiv:1705.02574 [astro-ph.CO]}
  \BibitemShut {NoStop}%
%%CITATION = ARXIV:1705.02574;%%
\bibitem [{\citenamefont {Okumura}\ \emph {et~al.}(2012)\citenamefont
  {Okumura}, \citenamefont {Seljak},\ and\ \citenamefont
  {Desjacques}}]{Okumura:2012xh}%
  \BibitemOpen
  \bibfield  {author} {\bibinfo {author} {\bibfnamefont {T.}~\bibnamefont
  {Okumura}}, \bibinfo {author} {\bibfnamefont {U.}~\bibnamefont {Seljak}}, \
  and\ \bibinfo {author} {\bibfnamefont {V.}~\bibnamefont {Desjacques}},\
  }\href {\doibase 10.1088/1475-7516/2012/11/014} {\bibfield  {journal}
  {\bibinfo  {journal} {JCAP}\ }\textbf {\bibinfo {volume} {1211}},\ \bibinfo
  {pages} {014} (\bibinfo {year} {2012})},\ \Eprint
  {http://arxiv.org/abs/1206.4070} {arXiv:1206.4070 [astro-ph.CO]} \BibitemShut
  {NoStop}%
%%CITATION = ARXIV:1206.4070;%%
\bibitem [{\citenamefont {Levi}\ \emph {et~al.}(2013)\citenamefont {Levi} \emph
  {et~al.}}]{Levi:2013gra}%
  \BibitemOpen
  \bibfield  {author} {\bibinfo {author} {\bibfnamefont {M.}~\bibnamefont
  {Levi}} \emph {et~al.} (\bibinfo {collaboration} {DESI}),\ }\href@noop {} {\
  (\bibinfo {year} {2013})},\ \Eprint {http://arxiv.org/abs/1308.0847}
  {arXiv:1308.0847 [astro-ph.CO]} \BibitemShut {NoStop}%
%%CITATION = ARXIV:1308.0847;%%
\bibitem [{\citenamefont {Amendola}\ \emph {et~al.}(2016)\citenamefont
  {Amendola} \emph {et~al.}}]{Amendola:2016saw}%
  \BibitemOpen
  \bibfield  {author} {\bibinfo {author} {\bibfnamefont {L.}~\bibnamefont
  {Amendola}} \emph {et~al.},\ }\href@noop {} {\  (\bibinfo {year} {2016})},\
  \Eprint {http://arxiv.org/abs/1606.00180} {arXiv:1606.00180 [astro-ph.CO]}
  \BibitemShut {NoStop}%
%%CITATION = ARXIV:1606.00180;%%
\bibitem [{\citenamefont {Bull}\ \emph {et~al.}(2016)\citenamefont {Bull} \emph
  {et~al.}}]{Bull:2015stt}%
  \BibitemOpen
  \bibfield  {author} {\bibinfo {author} {\bibfnamefont {P.}~\bibnamefont
  {Bull}} \emph {et~al.},\ }\href {\doibase 10.1016/j.dark.2016.02.001}
  {\bibfield  {journal} {\bibinfo  {journal} {Phys. Dark Univ.}\ }\textbf
  {\bibinfo {volume} {12}},\ \bibinfo {pages} {56} (\bibinfo {year} {2016})},\
  \Eprint {http://arxiv.org/abs/1512.05356} {arXiv:1512.05356 [astro-ph.CO]}
  \BibitemShut {NoStop}%
%%CITATION = ARXIV:1512.05356;%%
\end{thebibliography}%

\end{document}